\documentclass[twocolumn]{aastex62}
\usepackage{amsmath}

\received{2019 Feb 24}
\revised{2019 June 19}
\accepted{2019 July 1}
\shorttitle{Interrelationship between halo mass and group properties}
\shortauthors{Man et al.}

\begin{document}

\title{THE FUNDAMENTAL RELATION BETWEEN HALO MASS AND GALAXY GROUP PROPERTIES}

\correspondingauthor{Ying-Jie Peng}
\email{yjpeng@pku.edu.cn}

\author{Zhong-yi Man}
\affil{Department of Astronomy, School of Physics, Peking University, Beijing, 100871, China}
\affil{Kavli Institute for Astronomy and Astrophysics (KIAA), Peking University, Beijing, 100871, China}
\affil{Department of Astronomy, Yale University, New Haven, CT 06520, USA}
\author{Ying-jie Peng}
\affil{Kavli Institute for Astronomy and Astrophysics (KIAA), Peking University, Beijing, 100871, China}
\author{Jing-jing Shi}
\affil{Kavli Institute for Astronomy and Astrophysics (KIAA), Peking University, Beijing, 100871, China}
\author{Xu Kong}
\affil{Key Laboratory for Research in Galaxies and Cosmology, Department of Astronomy, University of Science and Technology of China, Hefei 230026, China}
\affil{School of Astronomy and Space Sciences, University of Science and Technology of China, Hefei 230026, China}
\author{Cheng-peng Zhang}
\affil{Department of Astronomy, School of Physics, Peking University, Beijing, 100871, China}
\affil{Kavli Institute for Astronomy and Astrophysics (KIAA), Peking University, Beijing, 100871, China}
\author{Jing Dou}
\affil{Department of Astronomy, School of Physics, Peking University, Beijing, 100871, China}
\affil{Kavli Institute for Astronomy and Astrophysics (KIAA), Peking University, Beijing, 100871, China}
\author{Ke-xin Guo}
\affil{Kavli Institute for Astronomy and Astrophysics (KIAA), Peking University, Beijing, 100871, China}
\affil{International Centre for Radio Astronomy Research, University of Western Australia, Crawley, WA 6009, Australia}

\begin{abstract}
We explore the interrelationships between the galaxy group halo mass and various observable group properties. We propose a simple scenario that describes the evolution of the central galaxies and their host dark matter halos. Star formation quenching is one key process in this scenario, which leads to the different assembly histories of blue groups (group with a blue central) and red groups (group with a red central). For blue groups, both the central galaxy and the halo continue to grow their mass. For red groups, the central galaxy has been quenched and its stellar mass remains about constant, while its halo continues to grow by merging smaller halos. From this simple scenario, we speculate about the driving properties that should strongly correlate with the group halo mass. We then apply the machine learning algorithm the Random Forest (RF) regressor to blue groups and red groups separately in the semianalytical model L-GALAXIES to explore these nonlinear multicorrelations and to verify the scenario as proposed above. Remarkably, the results given by the RF regressor are fully consistent with the prediction from our simple scenario and hence provide strong support for it. As a consequence, the group halo mass can be more accurately determined from observable galaxy properties by the RF regressor with a 50\% reduction in error. A halo mass more accurately determined in this way also enables more accurate investigations on the galaxy$-$halo connection and other important related issues, including galactic conformity and the effect of halo assembly bias on galaxy assembly.
\end{abstract}


\section{Introduction}
\label{sec_intro}
In the context of the $\Lambda$CDM paradigm, the formation and evolution history of galaxies are closely correlated with the hierarchical growth of the dark matter halos in which galaxies reside. Studying the interrelationships between various properties of galaxies and their host dark matter halos can help us better understand the galaxy formation physics, provide a basis for interpreting the large-scale structure observation, constrain the cosmological parameters, and distinguish various dark matter models to probe the nature of dark matter (see \citealt{Wech 18} for a comprehensive review on this topic). 

Currently, there are several ways to obtain halo mass in observation. In galaxy clusters, measuring the line-of-sight (LOS) velocity dispersion of the galaxies or the temperature and density of the hot intracluster medium can both directly derive the halo mass of the cluster through virial theorem and hydrostatic equilibrium, respectively. The abundance matching (AM) technique (e.g. \citealt{Kra 99, Mou 02, Tas 04, Vale 04, Yang 05, Con 06, Yang 07, Mo 10}), as a simple and powerful tool, provides an indirect way to derive the halo mass of galaxy groups. The key assumption of AM is that the most massive central galaxy lives in the most massive dark matter halo, followed by the second most massive central galaxy living in the next most massive halo, and so forth. Based on this, given a halo mass function, one can in principle assign halo mass to a central galaxy according to its stellar mass ranking. Weak gravitational lensing is another powerful tool for measuring the halo mass distribution (e.g. \citealt{Man 06}; \citealt{Luo 18}). 

Based on the assumption that the total stellar mass/luminosity of a galaxy group is correlated with its halo mass, \citet{Yang 05, Yang 07} assign a halo mass to each galaxy group identified by their group finder. As one of the most widely used Sloan Digital Sky Survey (SDSS) group catalogs, \citet{Yang 07} group catalogs have inspired a series of studies on correlations among galaxies, halos, and the large-scale environment (e.g. \citealt{Peng 10, Peng 12, Wang 13, Lac 14, Peng 14a, Bal 16, Spi 18, Wang 18, Gra 18, Dra 18}). Using the halo mass estimated from the AM technique, the signature of halo assembly bias can be detected \citep{Wang 08, Wang 13, Lac 14}. For instance, the low specific star formation rate (sSFR) sample is found to be clustered more than the high sSFR sample at a given halo mass. However, a signature of assembly bias was not found in \citet{Lin 16}. \citet{Lin 16} conclude that it is likely due to either the inaccurate mean relationship between total luminosity and halo mass, or the fact that the scatter in the estimated halo mass using the AM technique correlates with physical properties of the galaxies, such as sSFR and star formation history (SFH). Therefore, we may include galaxy observables (other than stellar mass) that may correlate with halo mass, and this could further minimize the scatter in the estimated halo mass, that is, allow us to derive a more accurate halo mass.

In this work, we will use machine learning (ML) techniques to analyze the underlying multicorrelations between various observable group properties and halo mass, and to predict the halo mass of galaxy groups. In recent years, ML has been used in predicting halo mass in several studies (\citealt{Nta 15, Nta 16, Nta 18, Arm 19, Ho 19, Cal 19}). For example, using the Support Distribution Machine, \citet{Nta 15, Nta 16} constrained the dynamical mass of galaxy clusters from distributions of LOS velocity dispersion of cluster members. The scatter in mass prediction is significantly reduced than in the $M-\sigma$ relation, even when clusters are contaminated with interloper galaxies. \citet{Arm 19} employed a variety of ML algorithms to predict cluster mass based on a set of dynamical observables other than LOS. Most recently, \citet{Ho 19} used a deep learning method, Convolutional Neural Networks, to produce dynamical mass estimates of galaxy clusters. 

However, most of these works are targeted on massive clusters and are mainly based on dynamical mass indicators, while our analysis is based on galaxy groups of a large mass range down to $\rm 10^{11} M_{\odot}/h$ with observables related to the SFH taken into account. We apply the Random Forest regressor (RF), a powerful supervised ML algorithm, to a subsample of the group catalog retrieved from the semianalytic model (SAM), L-GALAXIES \citep{Hen 15}. This model is built upon a cosmological $N$-body simulation, covering a relatively large mass range with good statistics, which makes it an ideal sample for our purpose. We expect that the blue and red groups following different stellar-halo mass relations (SHMR) may have different assembly histories. We will hence choose galaxy group observables based on the existing understanding of the galaxy$-$halo connection and perform the analysis separately for these two samples. We will use RF to identify the most important group properties in determining halo mass and look for analytical formulae of halo mass as a function of selected galaxy group properties. 

The layout of the paper is as follows. In section \ref{sec_2}, we will briefly introduce the SAM (L-GALAXIES), the mock catalog, and the galaxy group samples. In section \ref{sec_3}, we will discuss in detail on the different SHMRs for blue and red centrals and how we select galaxy group properties to be used in our analysis. In section \ref{sec_ML}, we will introduce the RF regressor and the configuration of the algorithm. The results will be presented in section \ref{sec_results}. We will summarize our main findings in section \ref{sec_sum}.

\begin{figure*}
\centering
\includegraphics[width=1\columnwidth]{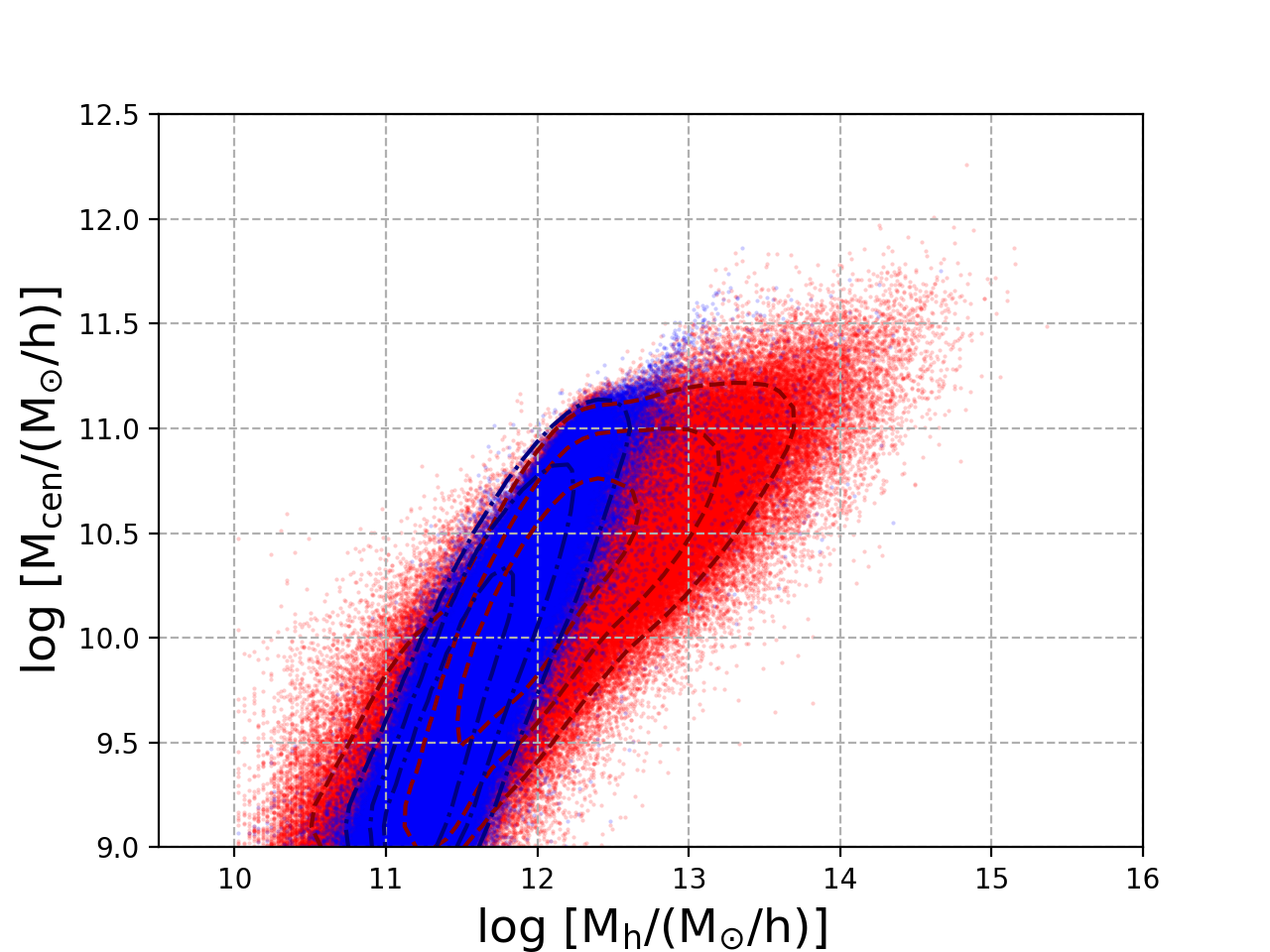}
\centering
\includegraphics[width=1\columnwidth]{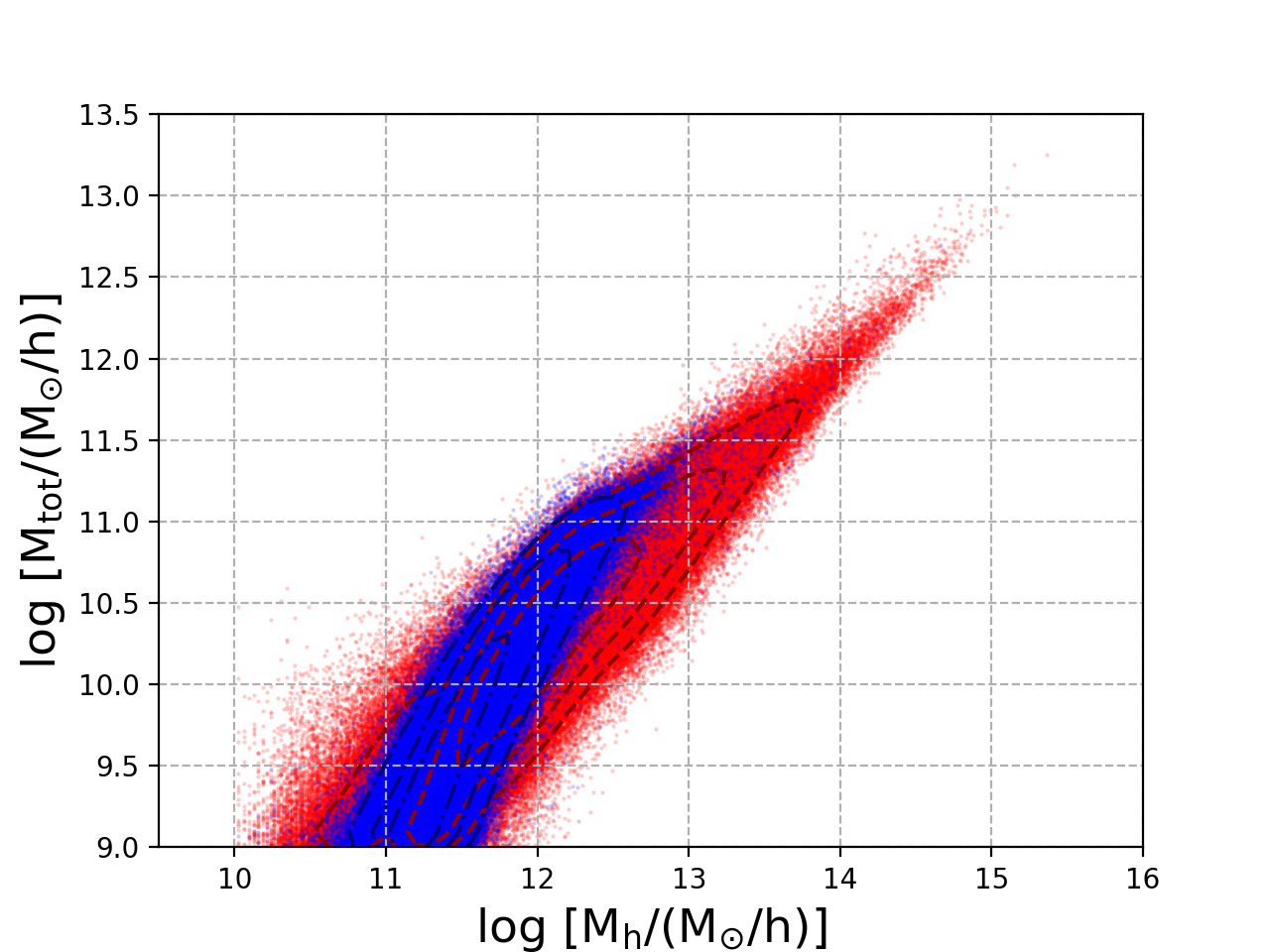}
\caption{Left panel: the stellar mass of central galaxies ($M_{\rm cen}$) as a function of halo mass ($M_{\rm h}$) for blue and red groups. Dashed-dotted and dashed contour lines show the number distributions of blue and red groups, respectively. Right panel: as in the left panel, but the y axis is the total stellar mass in the group ($M_{\rm tot}$). The correlation between halo mass and total stellar mass is tighter than the SHMR, implying the stellar mass of a galaxy group is a stronger indicator of halo mass. \label{fig_1}}
\end{figure*}

\section{Data}
\label{sec_2}
\subsection{Semianalytical Galaxy Formation Model}
\label{subsec_model}
Semianalytical model is a computationally efficient way to model the galaxy formation and evolution, by describing the various physical processes analytically and tracing the dark matter halo merger trees. In this work, we use the publicly released data of the latest Munich semianalytical model, L-GALAXIES\footnote{http://gavo.mpa-garching.mpg.de/MyMillennium/} \citep{Hen 15}. This model is built on the Millennium \citep{Spr 05} and Millennium-II \citep{Boy 09} simulations rescaled to the \textit{Planck} cosmology \citep{Planck 14}: $\sigma_8=0.829$, $H_0=67.3\ {\rm km} s^{-1} {\rm Mpc}^{-1}$, $\Omega_{\Lambda}=0.685$, $\Omega_m=0.315$, $\Omega_b=0.0487$ ($f_b=0.155$) and $n=0.96$. Compared with previous Munich galaxy formation models (e.g. \citealt{Guo 11, Guo 13}), the model has made several changes, such as delaying the reincorporation of wind ejecta, lowering the gas density threshold for star formation, modifying the radio-mode feedback, and eliminating ram-pressure stripping in halos smaller than $\sim$ $10^{14}M_\odot/h$ for satellites. Besides, L-GALAXIES has been careful with the observational errors: in the MCMC sampling of the model, \citet{Hen 13} used multiple ``good'' determinations of each observational property, took the scatter among them (together with the quoted statistical errors) to suggest likely systematic uncertainties.

The model employs the Markov Chain Monte Carlo (MCMC) method to adjust the parameter space to match observations. It has been carefully calibrated against observed stellar masses and passive fraction of galaxies within the redshift range of $0 < z < 3$, to produce the observed evolution of stellar mass function (SMF) and the distribution of sSFR. We are using the mocks produced from L-GALAXIES because SAMs like L-GALAXIES still produce more accurate stellar mass function than hydro-simulations, which indicates a more accurate SHMR. Also, SAMs usually have a much larger volume (500 Mpc/h for L-GALAXIES) than typical hydro-simulations, leading to larger training samples, better statistics and less cosmic variance.

\begin{figure*}
\centering
\includegraphics[width=1\columnwidth]{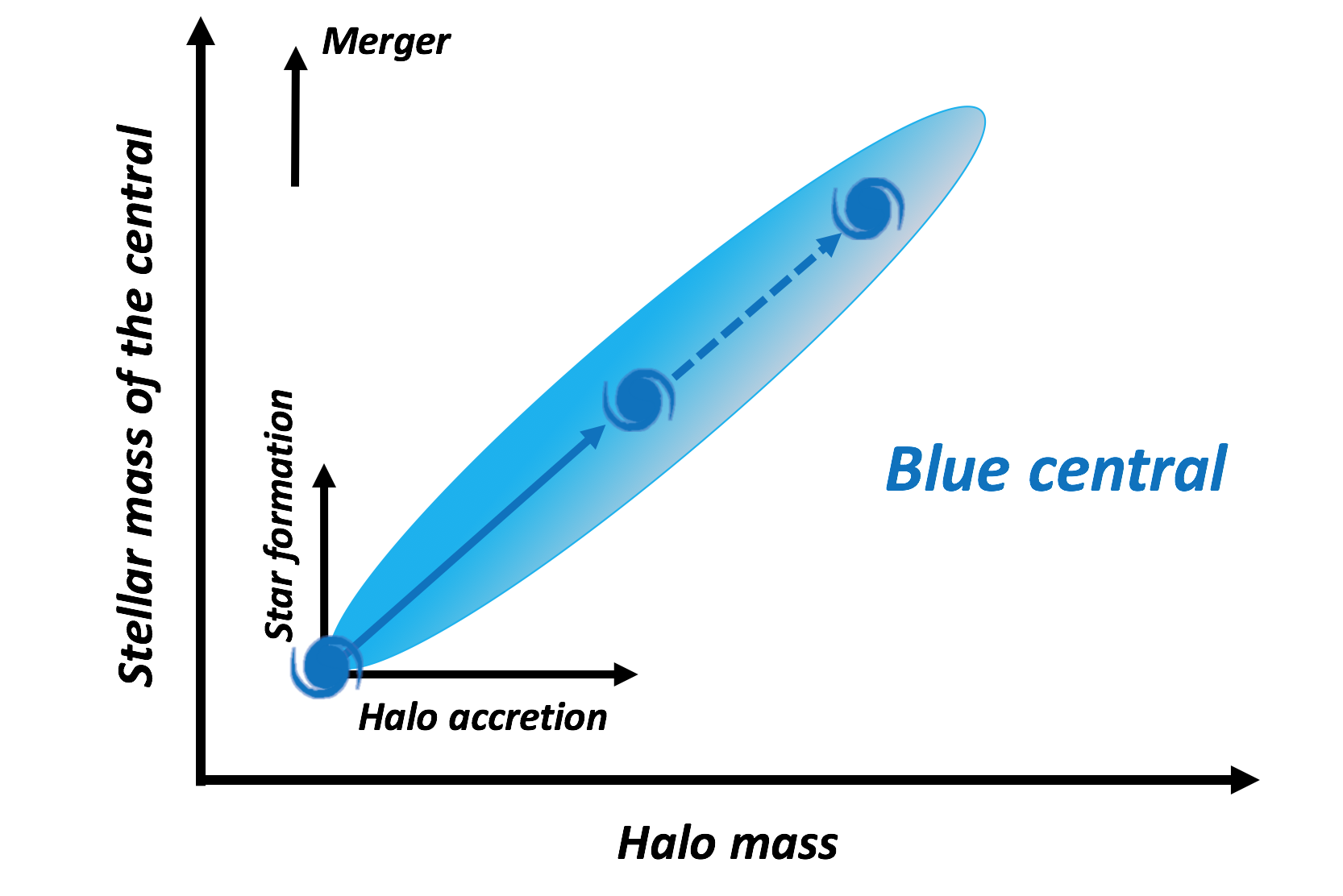}
\centering
\includegraphics[width=1\columnwidth]{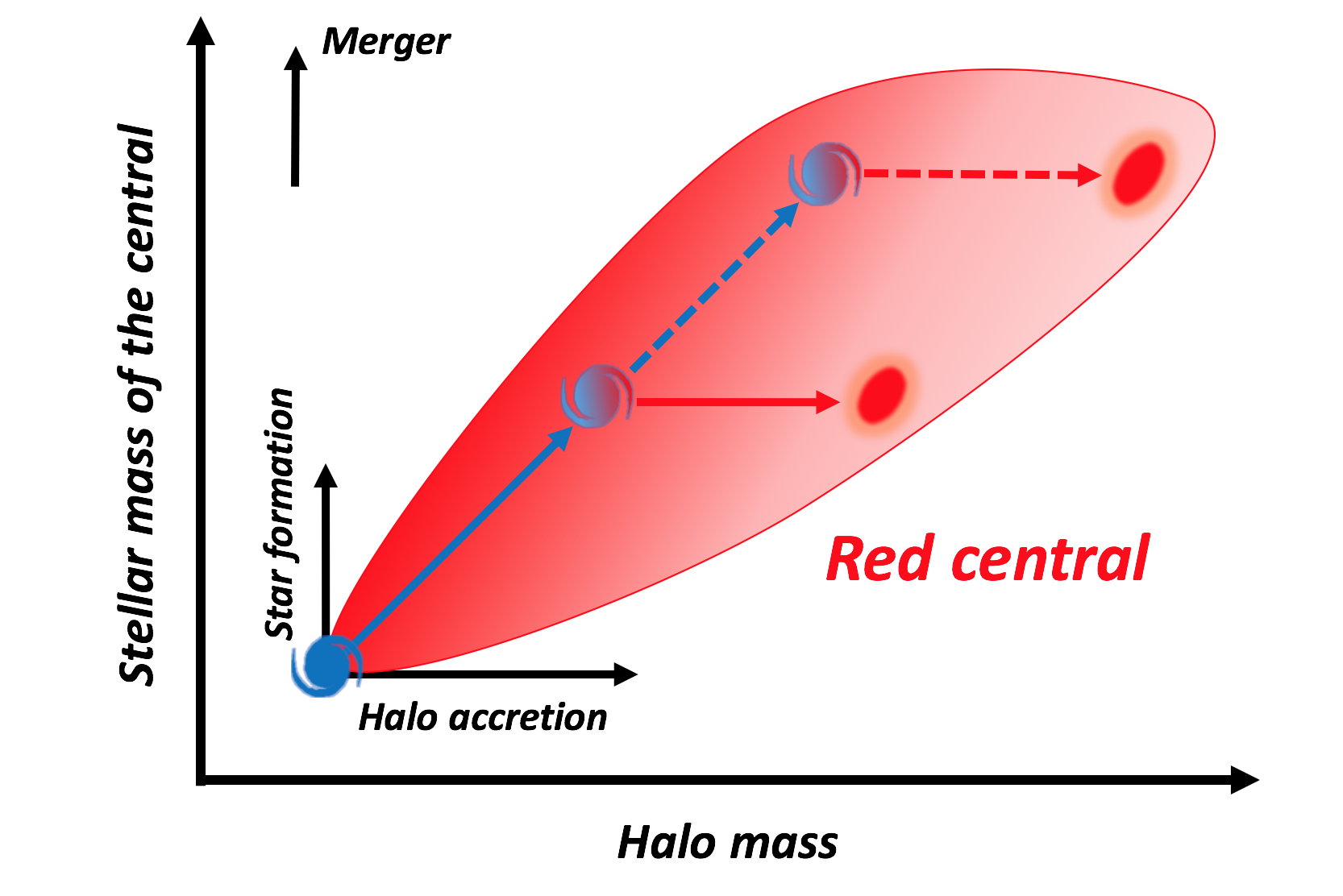}
\caption{Illustrations of different SHMRs for blue and red centrals. Left panel: For blue groups, because both the central galaxy and its host halo will continue to grow their masses simultaneously, the blue central galaxies are expected to move diagonally to the right and upward (indicated by the blue arrow). As a consequence of such coupled coevolution of the central and its host halo, a relatively tight relation between the stellar mass of the central and its host halo mass is expected. Right panel: For red groups, when the central is quenched at some point, its stellar mass remains about constant unless additional stellar mass is accreted through subsequent mergers, while its halo continues to grow by merging smaller halos. In other words, the growth of the halo has been decoupled from the growth of the central once the central was quenched, after which the red central has been moving horizontally to the right, as indicated by the red arrow. The triangular shading is due to the fact that not all of the quenched centrals will remain centrals during their horizontal evolution to the right (the red arrow). More massive red centrals will stand a higher chance of surviving as centrals following substantial growth in the mass of their parent halos. \label{fig_2}}
\end{figure*}

\subsection{Galaxy Group Samples}
\label{subsec_sample}
The aim of the work is to predict the halo mass of the galaxy groups in L-GALAXIES using the observable group properties. We select all galaxies with stellar mass above $10^{9}M_\odot/{h}$ in the snapshot with $z\sim0.05$ from the catalog based on the Millennium simulation. The final catalog consists of 3,632,259 galaxies, including 2,206,529 centrals and 1,425,730 satellites identified by the friends-of-friends group finder.

We separate galaxy groups into blue and red according to the color of the central galaxies. We adopt a widely used selection criteria (e.g. \citealt{Dar 16, Shen 17, Lai 18}) where the quiescent galaxies are defined by $NUV-r>3(r-J)+1$ and $NUV-r>3.1$ \citep{Ilb 09, Wil 09}. Both color are in the restframe with dust extinction included. $NUV-r$ is a good indicator of the current versus past star formation activity (\citealt{Mar 07, Arn 07}) and the two-color selection criteria can effectively differentiate between dusty star-forming galaxies and quiescent galaxies.

\section{Galaxy$-$halo connection \label{sec_3}}
\subsection{SHMR \label{subsec_SHMR}}
The SHMRs for blue and red groups in our mock catalog are shown in the left panel of Figure \ref{fig_1}. It is evident that the blue and red centrals follow different trends, as in previous works (e.g. \citealt{More 11, Peng 12, Wang 12, Rod 15, Man 16, Zu 16}). At a given halo mass, blue centrals on average have larger stellar masses than red centrals. At a fixed stellar mass of the centrals, the red centrals are living, on average, in more massive halos with more satellites than the blue ones, and the red fraction of centrals increases with halo mass. As discussed in \citet{Peng 12} and \citet{Peng 14b}, these results can be explained by a simple scenario in which quenching is a result of stellar mass alone, as illustrated in Figure \ref{fig_2} below.

The cartoons in Figure \ref{fig_2} illustrate the different SHMRs for blue and red centrals. For blue groups (left panel), the central galaxies grow their stellar mass by star formation or mergers. Meanwhile, their host dark matter halos also continue to grow by merging with other halos. Therefore, the blue central galaxies are expected to move diagonally to the right and upward (indicated by the blue arrow). As a consequence of such coupled coevolution of the central and its host halo, a relatively tight relation between the stellar mass of the central and its host halo mass is expected. 

For red groups, when the central is quenched at an early epoch (by certain physical mechanisms), its stellar mass remains about constant unless additional stellar mass is accreted through subsequent mergers, while its halo continues to grow by merging smaller halos, irrespective of the star formation status of the central galaxy. In other words, the growth of the halo has been decoupled from the growth of the central once the central was quenched, after which the red central has been moving horizontally to the right, as indicated by the red arrow. The triangular shading is due to the fact that not all of the quenched centrals will remain centrals during their horizontal evolution to the right (as marked by the red arrow). More massive red centrals will stand a higher chance of surviving as centrals following substantial growth in the mass of their parent halos, while less massive red centrals may become satellites, or may even disappear completely if they merge with larger galaxies, when their parent halos merge with more massive halos (with more massive galaxies). In both cases, they will no longer exist on this plot. Consequently, there are fewer low-mass red centrals that continue to evolve to the right than high-mass red centrals, which leads to the triangular red shading in the right panel of Figure \ref{fig_1}.

Putting this all together, first of all, we see that the total stellar mass of a group is expected to correlate strongly with the group halo mass, since the total stellar mass is the best indicator of both the overall SFH of all group members and the merging history of the halo. Second, as above, the different assembly histories for blue and red groups can produce different SHMRs. For blue groups, in addition to the total stellar mass of the
group, the properties that indicate the SFH of the centrals, such as star formation rate and color of the centrals, should also correlate strongly with group halo mass. For red groups, in addition to the total stellar mass of the group, the properties that can indicate the quenching epoch of the centrals (e.g., stellar age of the central) and the halo growth history (e.g., group richness) should also correlate strongly with group halo mass.

To investigate the complicated nonlinear and nonorthogonal multicorrelations between various observable galaxy properties and group halo mass, and to verify the simple scenario above, we employ ML techniques. As above, we will treat the blue groups and red groups separately in our following analysis. This is hence different from many previous studies (\citealt{Nta 15, Nta 16, Arm 19, Cal 19}), which do not differentiate between blue groups and red groups.

We stress that our primary goal is to use the ML technique to verify the scenario as proposed above, which gives a simple description of the different coevolution histories for blue and red groups with their dark matter halos. Since the ML algorithms can also quantify the correlation between various observable galaxy properties and group halo mass, in return, the group halo mass can be more accurately predicted from observable galaxy properties.

\subsection{Galaxy Group Properties}
\label{subsec_group_properties}

Figure \ref{fig_1} shows the correlation between group halo mass and the stellar mass of the central in the left panel and group halo mass and total stellar mass of the group in the right panel. As in the analysis in the previous section, the correlation in the right panel is evidently tighter than that in the left panel, indicating that the total stellar mass of the group plays a critical role in determining the group halo mass. Meanwhile, the scatter becomes progressively larger toward the low-mass end, which implies that other group properties may start to become important in determining the group halo mass in the low-mass regime.

As discussed in the previous section, galaxy group properties related to the assembly formation history of the centrals and their host halos should directly contribute to halo mass. We also include other key observable properties in our analysis as follows:

\begin{itemize}
    \item Stellar mass of the central galaxy ($M_{\rm cen}$), used in the standard AM approach.
    \item Total stellar mass in the group ($M_{\rm tot}$), used in the AM of \citet{Yang 05, Yang 07}.
    \item Group richness, defined as the total number of group members above a certain mass threshold (see \citealt{Kno 09}). As discussed in the previous section, it is expected to correlate with halo mass for characterizing the halo growth history. In addition, \citet{Peng 12} used richness to study the environmental effect of galaxy quenching and found it a good proxy of halo mass on group scales. 
    \item SFR of the central galaxy (SFR). As discussed in the previous section, SFR and color can reveal the SFH of the central, thus correlating with group halo mass.
    \item $NUV-r$ color of the central galaxy ($NUV-r$), which is a good indicator of the current versus past star formation activity (\citealt{Arn 07, Mar 07})
    \item $r$ band weighted stellar age of the central galaxy (Age). As discussed in the previous section, the stellar age of the central can indicate the quenching epoch of the red centrals, and may hence be correlated with halo mass \citep{Lac 11}. The stellar age is also used to investigate the galaxy assembly bias in SDSS \citep{Lac 14}. 
    \item Compactness of the group in terms of the projected median distance of all satellites to the central galaxy ($\langle D_{\rm sat}\rangle$). This might be taken as an observational manifestation of halo concentration, which is a promising secondary parameter of dark matter halos driving the galaxy$-$halo connection \citep{Wech 06, Fal 10}.
    \item Luminosity gap in the $r$ band, defined as the $r$-band luminosity difference between the brightest and the second brightest galaxies within the group ($\Delta m_{\rm r}$). It has been taken as a secondary halo mass indicator besides luminosity or stellar mass of the central galaxy (\citealt{More 12, Hea 13, Shen 14, Lu 15}).  
    \item Bulge$-$total mass ratio of the central galaxy (B/T). It is found to be correlated with the SFH of the galaxies \citep{Che 12, Wak 12, Blu 14}, which have been available for SDSS \citep{Sim 11}.
   
\end{itemize}

The virial mass of the halo where the galaxy group lies is defined as the dark matter mass enclosed in the spherical volume within which the average density is 200$\rho_{\rm crit}$, with $\rho_{\rm crit}$ being the critical density of the universe. 

The L-GALAXIES model has been calibrated to reproduce the real values of group properties (section \ref{subsec_model}) that are basically available in observation. In particular, L-GALAXIES gives so far one of the most accurate fits of the SMF in the local universe, including SDSS \citep{Bal 08, Li 09} and the Galaxy And Mass Assembly (GAMA) survey \citep{Bal 12}. The galaxy group and group richness in L-GALAXIES are derived via the friends-of-friends method, similar to those used in observation (e.g. \citealt{Yang 05, Ber 06, Yang 07}). Compared to previous models (e.g. \citealt{Guo 13, Hen 13}), the L-GALAXIES \citep{Hen 15} also has improvements in matching the distributions of color, SFR \citep{Bri 04, Sal 07}, and stellar age \citep{Gal 05} at a fixed stellar mass. 

However, to a large or small extent, there is always systematic difference in the absolute mean values (e.g. stellar mass, SFR, color and etc.) between model predictions and observations, as is the same case with the scatters around the mean value. In order to reduce the potential bias introduced by the systematic difference in both the mean value and scatter around the mean, for a given galaxy group property ($P$), we use its renormalized dimensionless forms $\rm P_n$, defined as

\begin{equation}
     \rm P_n=\frac{ P- \bar{P}}{\sigma_P}\label{equ_1}
\end{equation}
where $\bar{P}$ and $\sigma_P$ are the mean and standard deviation of the parameter $P$.

By using this renormalized dimensionless form, we have assumed $P$ follows a normal distribution in both SAM and observations, and the difference between the two distributions depends only on these two parameters. While this assumption is not perfectly satisfied (e.g. for stellar age), the form we use could still reduce the biases from the offsets between SAM and observations in either the mean value or the scatter around the mean. In practice, we apply equation \ref{equ_1} to SFR, $NUV-r$, age, $\Delta m_{r}$ and B/T ratio, whose absolute values are more likely to deviate from observation, while keeping the original values for $M_{\rm cen}$, $M_{\rm tot}$, richness, and $\langle D_{\rm sat}\rangle$ given that the SMF and galaxy clustering are the most accurately predicted properties in SAMs. The following analyses are hence based on these renormalized input group observables.  

\begin{table*}
\begin{center}
\caption{Number of Galaxies in the Training, Validation, and Test Subsets for Blue and Red Groups.}
\begin{tabular}{crrrrrrrrrr}
\tableline\tableline
Color  & $N_{\rm train}(80\%)$ & $N_{\rm validation}(10\%)$ & MSE$_{\rm validation}$ & $N_{\rm test}(10\%)$ & MSE$_{\rm test}$ & $\rho_{\rm test}$ \\\tableline
Blue & 1305238  & 163154 & 0.003454 & 163154 & 0.003494 & 0.976\\
Red & 459986  & 57498 & 0.0231 & 57498 & 0.02362 & 0.966\\
\tableline
\end{tabular}
\tablecomments{The MSEs in the validation and test subsets using the optimal RF regressor are listed. The Pearson correlation parameter $\mathbf{\rho}$ is used to quantify the linear correlation between the predicted and true halo masses in the test sample. \label{tbl-1}}
\end{center}
\end{table*}

\begin{figure*}
\centering
\includegraphics[width=0.8\columnwidth]{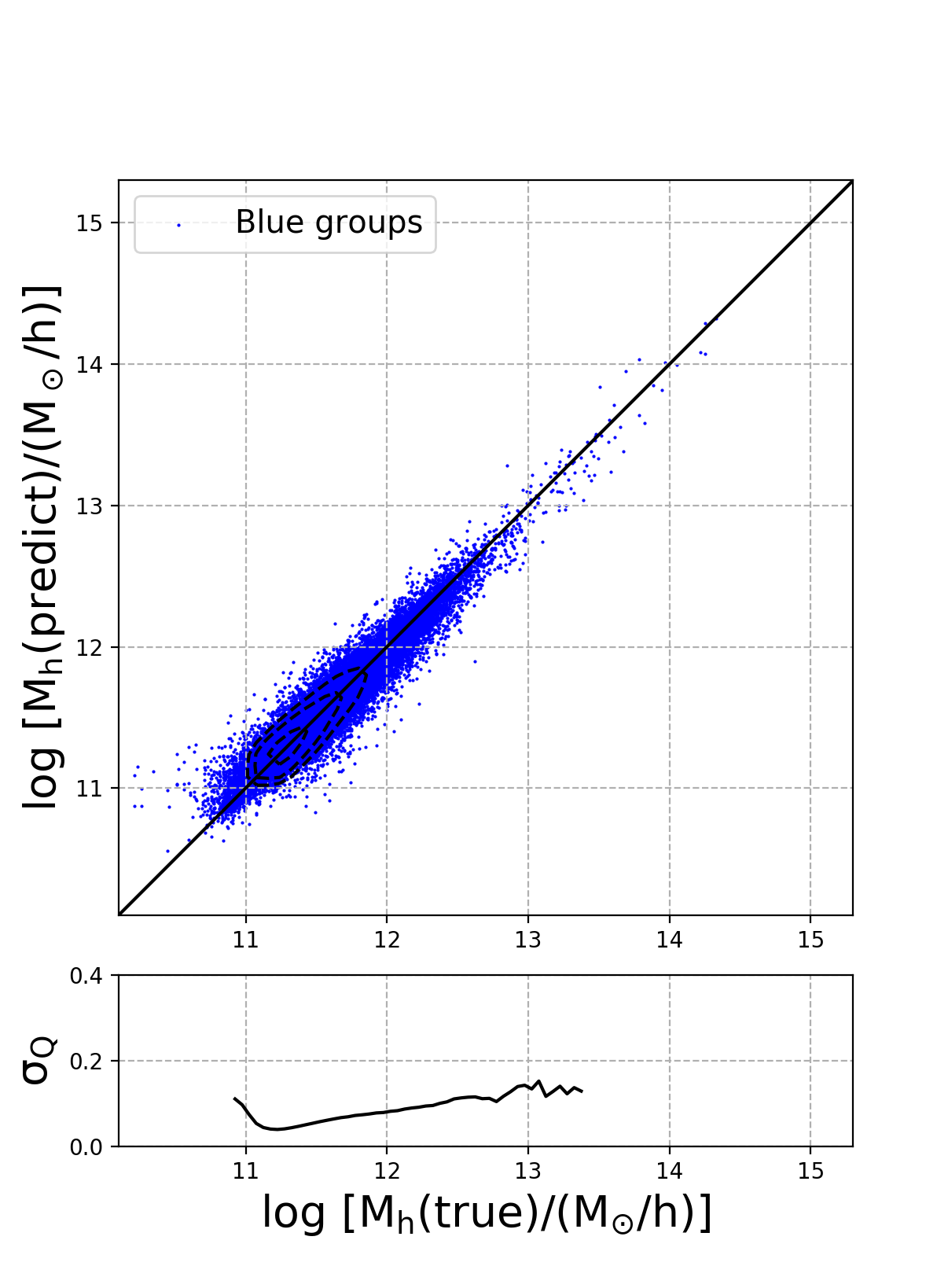}
\centering
\includegraphics[width=0.8\columnwidth]{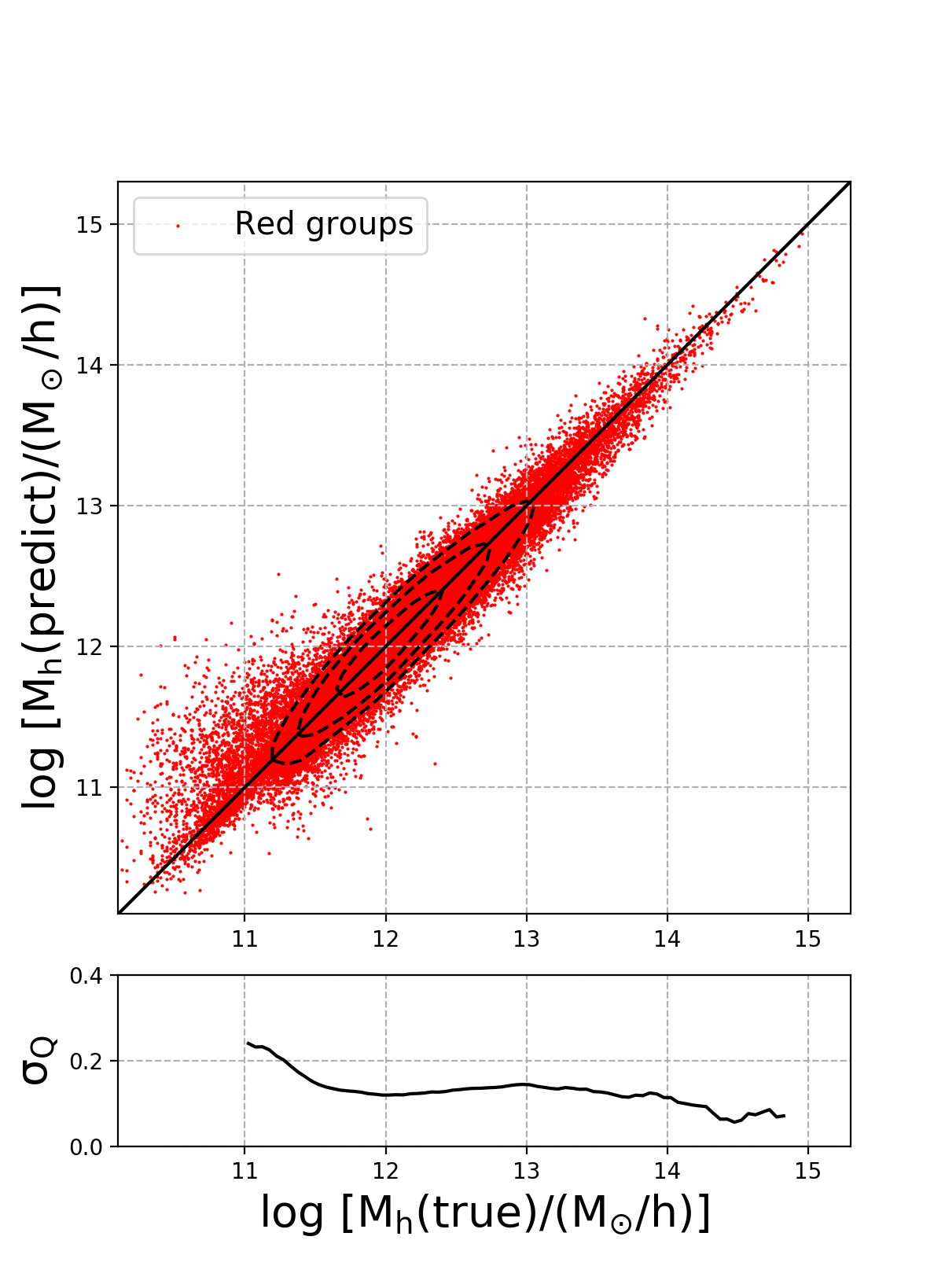}
\caption{Predicted halo mass as a function of the true halo mass in the test sample, for blue (left) and red (right) groups. Dashed contour lines in each panel show the number distributions of groups. Bottom panels show $\sigma_Q$ as a function of true halo mass, with a value around 0.1 dex. The MSE is 0.003494 for blue groups and 0.02362 for red groups. The RF regressor in general recovers the halo mass rather well. The 1:1 diagonal lines go through the ridge of all contour lines, indicating that there is little systematic difference between the predicted value and the true value of the halo mass.\label{fig_3}}
\end{figure*}

\section{Machine Learning}
\label{sec_ML}

\subsection{Random Forest Regressor \label{subsec_RF}}
As for the algorithm of ML, we adopt the RF (\citealt{Bre 01}) regressor in Python library \emph{scikit-learn} \footnote{http://scikit-learn.org} \citep{Ped 11}. The RF algorithm is highly efficient, easy to use, and capable of dealing with multifeature data without requiring feature selection. The unit of RF is the decision tree, a tree-like model of decisions. The root node of the tree is split into different decision nodes, which are further split into more nodes. A final node that does not split anymore is a leaf (decision). An RF is a combination of randomly generated decision trees from the same data set. Each tree is trained individually using a random subset of features, which can thus mitigate the problem of overfitting. Using RF can increase the
signal-to-noise ratio of the prediction since errors across different trees are likely to cancel each other out. Considering the interrelationships between galaxy observables and halo mass are mostly nonlinear, RF could possibly yield better a prediction than the linear modeling methods, such as ordinary least squares and partial least squares \citep{Mar 17}.

\begin{figure*}
\centering
\includegraphics[width=0.8\columnwidth]{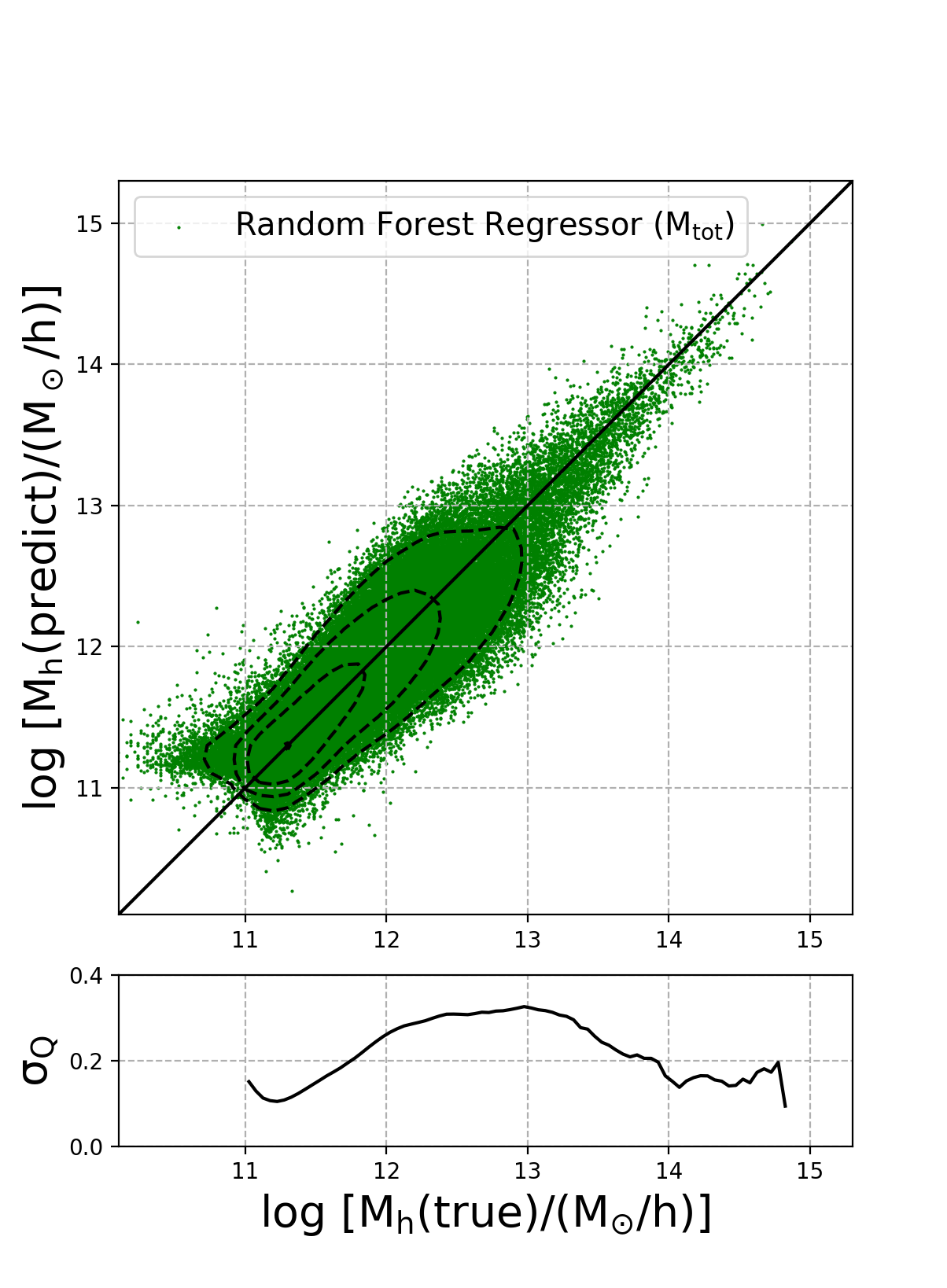}
\centering
\includegraphics[width=0.8\columnwidth]{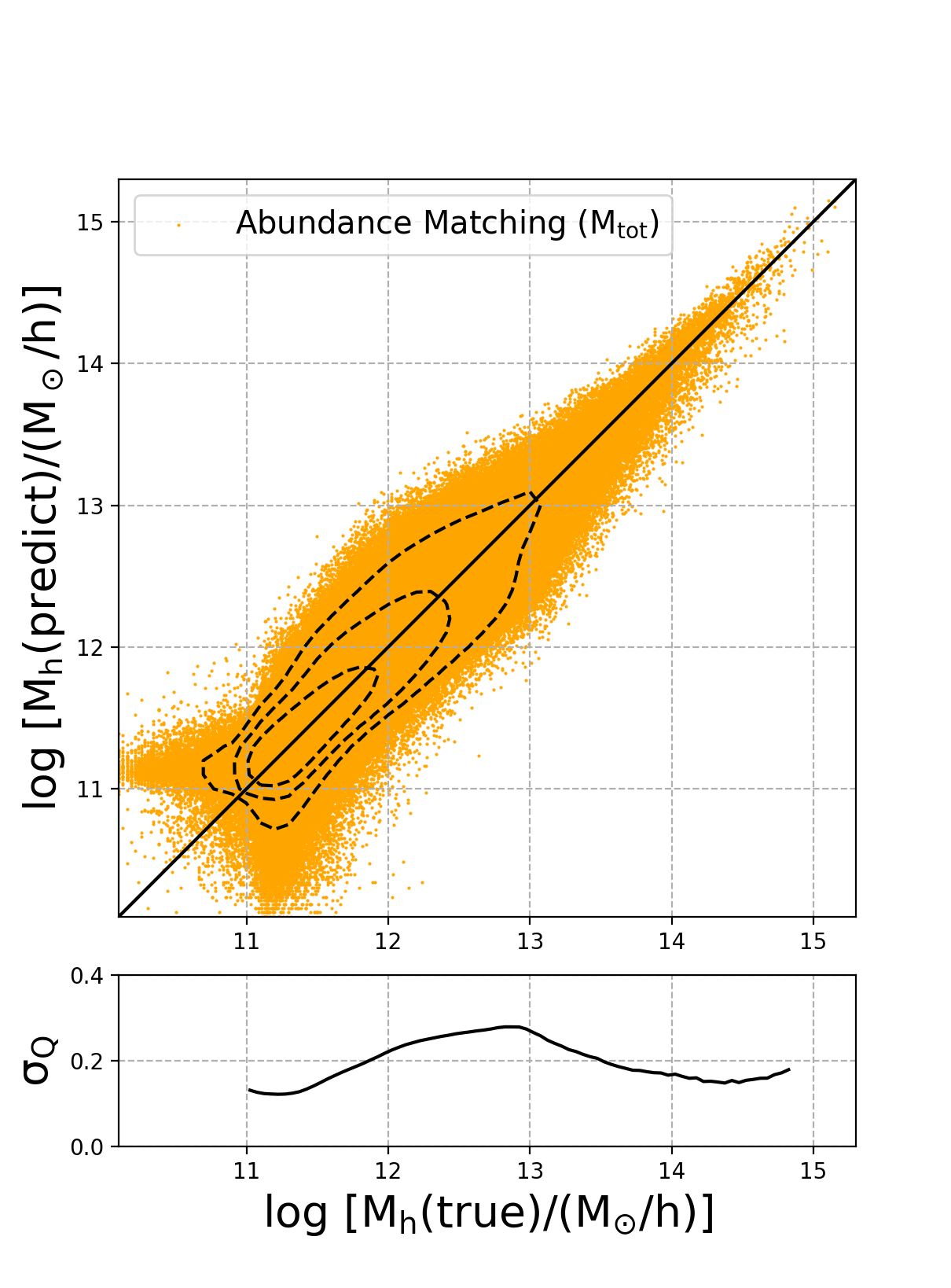}
\caption{Predicted halo mass as a function of the true halo mass using only the group total stellar mass ($M_{\rm tot}$) for groups regardless of the group color. The left panel shows the halo mass produced by the RF regressor in the test sample. The right panel shows the derived halo mass of the full sample by using the total stellar mass AM approach in \citet{Yang 07}. Dashed contour lines in each panel show the number distributions of groups. The bottom panels show $\sigma_Q$ as a function of true halo mass, with a value around 0.2 dex. The MSE is 0.04193 for RF regressor and 0.03241 for AM. The results are similar to each other, suggesting the RF regressor seems not superior to AM when using $M_{\rm tot}$ as the only input feature. \label{fig_4}}
\end{figure*}

The RF can make accurate predictions after being well trained, but it is also important to interpret the ``black-box-like'' model to understand the contribution of each input parameter to any particular predictions. Feature importance is a popular approach to quantifying the contribution of different input features to the predicted variable (halo mass in our case). There are different ways to compute the feature importance. For instance, \citet{Pal 13} proposed three methods: median, cluster analysis, and log-likelihood. In \emph{scikit-learn}, RF regressor ranks the relative importance of the input features based on the \emph{gini importance} \citep{Ped 11}, which is the most common approach. The \emph{gini importance} of a feature is computed by measuring its efficiency in reducing variance when creating decision trees within RF regressor. The feature importance given in \emph{scikit-learn} will provide a clear view of the prediction from the model and can be used to explore the correlations between halo mass and different group properties.

\subsection{Training, Validation, and Test Samples\label{subsec_VT}}
We adjust the hyperparameters of RF regressor to enhance its efficiency and accuracy. For instance, \texttt{n\underline{ }estimators} is the number of trees in the forest. A larger value could lead to a better prediction but with a higher cost of time. \texttt{max\underline{ }features} is the size of the random subsets of features to consider when splitting a node. Increasing the size can reduce the variance, but also increase the bias. Other adjustable hyperparameters include the maximum depth of trees, minimum features when splitting a node, and the minimum size of a leaf. Although some of the hyperparameters have empirically optimal values, one still needs to cross-validate different combinations of them to enhance the performance of RF regressor with reasonable computation time. 

In order to predict the halo mass from given galaxy group properties via RF regressor, we randomly divide the blue and red group samples into training, validation, and test sets, as shown in Table \ref{tbl-1}. The training sample occupies 80\% of the full sample, while the validation and test samples each occupies 10\%. The training sample is used to train the RF, and the validation sample is used to tune the hyperparameter space of the RF routines that minimize the mean square errors (MSE) between the true halo mass ($M_{\rm h, true}$) and the halo mass predicted by the RF model ($M_{\rm h, predict}$). The MSE is defined as

\begin{equation}
    \rm MSE=\frac{1}{N}\sum_{i=1}^{N}(M_{\rm h,predict}^i-M_{\rm h,true}^i)^2.
\end{equation}

After we obtain the optimal set of hyperparameters for RF regressor, we apply the regressor to the test sample, which is used to evaluate the performance of the trained algorithm. The MSEs of the test samples and the Pearson correlation factors of true and predicted halo mass are shown in Table \ref{tbl-1}. It is evident that the MSEs of the test sets (that have not been used in model training and validation) are identical to the validation sets.

\begin{table*}
\begin{center}
\caption{The Relative Importance of All Input Parameters Employed for the Training of the RF Regressor and the OOB Scores for the RF Algorithm}
\begin{tabular}{crrrrrrrrrr}
\tableline\tableline
Groups  & $M_{\rm cen}$ & $M_{\rm tot}$ & Richness & $\rm SFR_n$ & $(NUV-r)_n$ & $\rm Age_n$ & $\langle D_{\rm sat}\rangle$ & $(\Delta m_{\rm r})_n$ & $(B/T)_n$ & OOB Score\\
\tableline
Blue  &  0.821\% & \textbf{76.930\%}  & 0.220\% & \textbf{15.674\%}  & \textbf{2.471\%} & 1.630\% & 0.399\% & 0.214\% & 1.641\% & 95.165\%\\
Red & 1.191\% & \textbf{65.700\%} & \textbf{16.800\%} & 1.564\% & 1.743\% & \textbf{7.763\%} & 0.731\% & 0.585\% & 3.921\% & 93.089\% \\
\tableline
\end{tabular}
\tablecomments{The definition for each input parameter is given in Section \ref{subsec_group_properties}. Observables with subscript $n$ are the renormalized dimensionless parameters. The OOB scores quantify the generalization accuracy of the regressor. The most important input parameter is the total stellar mass for both group samples, and the second and third most important parameters are SFR and NUV-r color of the central galaxy for blue groups, and group richness and stellar age of the central galaxy for red groups (in bold). \label{tbl-2}}
\end{center}
\end{table*}

\section{Results}
\label{sec_results}
\subsection{Predicting Halo Mass}
The halo masses are predicted for the test samples by using the optimal RF regressor tuned based on the validation samples. We compare the predicted halo mass with the true halo mass in Figure \ref{fig_3} for blue groups (left panel) and red groups (right panel). The halo masses are recovered rather well via RF regressor: the 1:1 diagonal line in both panels (left for blue groups and right for red groups) goes through the ridge of all contour lines, indicating that there is little systematic difference between the predicted value and the true value of the halo mass. To quantify the distance to the 1:1 diagonal line ($M_{\rm h, true}=M_{\rm h, predict}$) for each point, we adopt the parameter $\sigma_Q$, as defined in \citet{Yang 07}:

\begin{equation}
\rm Q\equiv \frac{1}{\sqrt{2}}({\rm log}\ M_{\rm h,  predict}-{\rm log}\ M_{\rm h, true})
\end{equation}
where $\sigma_Q$ is the standard deviation of $Q$ at a given ${\rm log}\ M_{\rm h, true}$

In the bottom panel of Figure \ref{fig_3}, on average $\sigma_Q$ is as small as 0.1 dex for blue groups, and it becomes slightly larger for red groups but is still less than 0.2 dex, except at the very low mass end. The MSE is only 0.003494 for blue groups and 0.02362 for red groups. In the traditional AM approach in which only the total stellar mass (or total luminosity) of the group is used to derive the halo mass, $\sigma_Q$ is $\sim$ 0.2 dex or up to 0.3 dex (e.g. \citealt{Yang 07}). The group total stellar mass is the only halo mass indicator employed in \citet{Yang 07}, where they use the AM method to match halo mass with $M_{\rm tot}$. To make a more direct comparison, we repeat our analysis by using $M_{\rm tot}$ as the only input parameter for the RF regressor. The results are shown in the left panel of Figure \ref{fig_4}. In this analysis, as in the usual AM approach, we have not differentiated between blue and red groups. We also apply the usual AM approach to our full sample by ranking and then assigning the halo mass to the group according to $M_{\rm tot}$ only. The results are shown in the right panel of Figure \ref{fig_4}. The derived $\sigma_Q$ values in both panels are identical, around $\sim$ 0.2 dex or up to 0.3 dex, which agrees well with \citet{Yang 07}. The MSE is 0.04193 for RF regressor and 0.03241 for AM, both of which are higher than the RF regressor with more input observables. It is interesting to note that the results shown in the two panels are very similar, indicating that the performance of the RF regressor is not superior to the usual AM approach when $M_{\rm tot}$ is used as the only input parameter. This is expected beacuse the advantage of RF is in exploring the complicated correlation between multiple input parameters. Compared to the scatters in Figure \ref{fig_3}, where a set of galaxy group properties have been used to predict the halo mass, the RF regressor evidently produces a more accurate prediction of the halo mass than the usual AM approach, reducing the scatter by about 50\%.

\subsection{Importance of Group Properties}
\label{subsec_feature_importance}
As discussed in section \ref{subsec_RF}, one of the great features of the RF regressor is that it can calculate and rank the relative importance of each input parameter in determining the output. Table \ref{tbl-2} shows the relative importance of each group property by using the training sample with our optimized RF regressor. Group properties with subscript ``n'' are the renormalized dimensionless parameters. The out-of-bag (OOB) scores in the last column characterize the overall accuracy of the regressor. The values for blue groups and red groups are both above 90\%, suggesting that the model predictions are quite accurate and sample-independent. 

Table \ref{tbl-2} shows that the total stellar mass ($M_{\rm tot}$) is the driving input parameter for both blue and red groups ($\sim70\%$), while the stellar mass of the central ($M_{\rm cen}$) is trivial ($\sim1\%$). This is because the information carried by $M_{\rm cen}$ is already contained in $M_{tot}$ which further includes an additional mass contribution from satellites. As discussed in section \ref{subsec_SHMR}, this is expected as the total stellar mass is the best indicator of both the overall SFH of all group members and the merging history of the halo. The $M_{\rm tot}-M_{\rm h}$ relation as shown in the right panel of Figure \ref{fig_1} apparently has a smaller scatter than other relations, for instance, the $M_{\rm cen}-M_{\rm h}$ relation shown in the left panel of Figure \ref{fig_1}. This is consistent with the basic assumption of the usual AM approach, where the halo mass is matched with the total stellar mass in the group only.  However, the scatter of the $M_{\rm tot}-M_{\rm h}$ relation is still significant, especially in the low-mass end. As in Table \ref{tbl-2}, the relative importance of the other parameters counts about 30\% in total, which has been missed in the usual AM approach. Therefore, bringing in additional information (i.e. other galaxy properties) will produce a more accurate halo mass.

It is interesting to note that the second and the third most important parameters for blue and red groups are different. For blue groups, properties that characterize the star formation activity of the central galaxies (SFR and $\rm NUV-r$) are most important next to the total stellar mass, while group richness and stellar age of the central galaxies are more important for red groups. This color dichotomy supports our scenario illustrated in section \ref{subsec_group_properties} that the blue and red centrals have distinct evolutionary histories. It is therefore necessary to make predictions separately for blue and red groups. 

Remarkably, these results are fully consistent with the prediction from the simple scenario as discussed in section \ref{subsec_SHMR} and illustrated in Figure \ref{fig_2}. 

It becomes clear from the analysis above that one key to improving the prediction of the halo mass is to differentiating between blue and red groups. In other words, the key is to add quenching in the analysis. As discussed earlier, before quenching happens, we have the simple coupled coevolution between the blue centrals and their dark matter halos. When the central is quenched, its stellar mass remains largely constant unless additional stellar mass is accreted through subsequent mergers (and passive evolution of the stellar population), while its dark matter halo continues to grow regardless of the star formation status of the central, that is, the growth of the halo is decoupled from the growth of the central. 

One may wonder why the blue and red groups have to be classified according to the color$-$color diagram, now that the colors of the centrals are already used as input parameters in the RF regressor. The reason is that although RF regressor is a powerful ML algorithm, it may not always successfully identify the two distinct evolutionary paths as discussed above. In other words, it may not capture the quenching process by itself. Therefore, if we take quenching into account by differentiating between blue and red groups when performing RF regressor, we can presumably improve its performance and produce a more accurate prediction of the halo mass. Another potentially important hidden process is merger. If we could quantify the composition of stellar mass for a given central galaxy (e.g., how much of its stellar mass is from in situ star formation and
how much is from mergers), we should be able to further improve the accuracy of the halo mass prediction by including it as a new input parameter. We will explore this in our subsequent work.

The analysis and discussion above also imply that a better understanding of the underlying physics, in particular these hidden correlations between multiple variables, will help to improve the performance of of the ML algorithms. Further evidence is found in applying the ML technique to predict photometric redshifts from multiband photometry data (e.g. \citealt{Kind 13}). Using color (i.e. the difference between the magnitudes in different bands) usually produces more accurate redshifts than using magnitudes directly in the photometric redshift estimation.

\subsection{Empirical Formulae}  
\label{subsec_empirical_formulae}
The analysis demonstrates the key relations between halo mass and galaxy group properties, which can be used to make a more accurate prediction of the group halo mass than the traditional AM approach. However, the RF regressor gives no explicit form of such underlying relations. In practice, analytical formulae will be more convenient and useful for halo mass prediction.

We use the ordinary least squares (OLS) regression model to empirically fit the true halo mass with the three most important parameters, as discussed above for blue and red groups. We use total stellar mass ($M_{\rm tot}$), renormalized $\rm SFR_n$ and renormalized $\rm (NUV-r)_n$ for blue groups, and we use total stellar mass, group richness, and renormalized stellar age ($\rm Age_n$) for red groups. 

To obtain the optimal fitting parameters, we use the full sample to fit our regression model for blue groups and red groups. Given the negative slope of the mass function of the central galaxies, most central galaxies are distributed in $10^{10}<M_\star<10^{10.5}M_{\odot}/h$ (see Figure \ref{fig_1}). To obtain a relation that would work equally well across the entire explored mass range, we randomly choose an equal amount of central galaxies in each stellar mass bin to generate a new random sample out of the original training sample. For blue groups, we end up having 126,000 galaxies in three stellar mass bins ranging from $10^{9}M_{\odot}/h$ to $10^{11.4}M_{\odot}/h$. For red groups, we have 159,000 galaxies in three stellar mass bins ranging from $10^{9}M_{\odot}/h$ to $10^{11.7}M_{\odot}/h$. 

\begin{figure*}
\centering
\includegraphics[width=1\columnwidth]{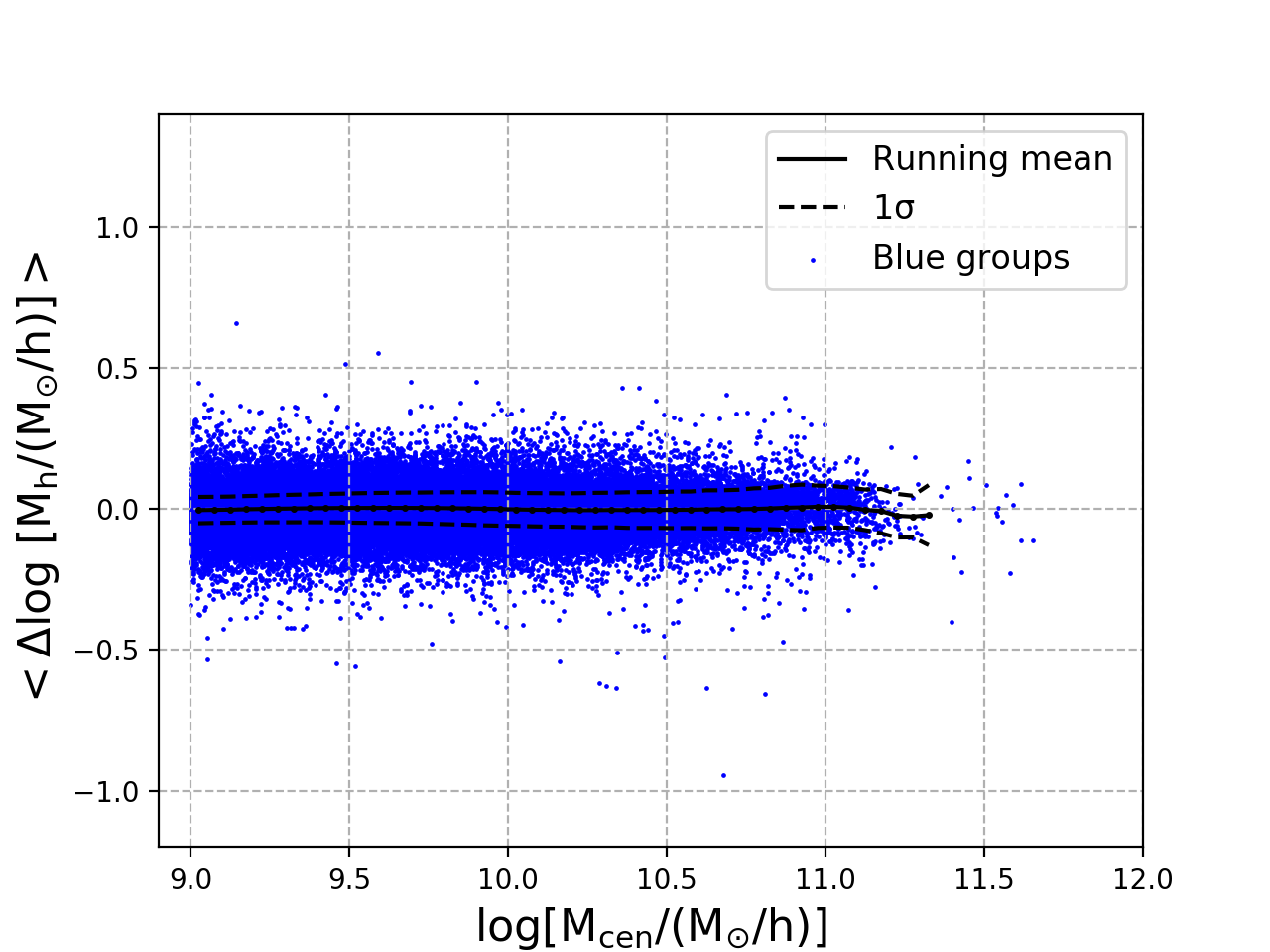}
\centering
\includegraphics[width=1\columnwidth]{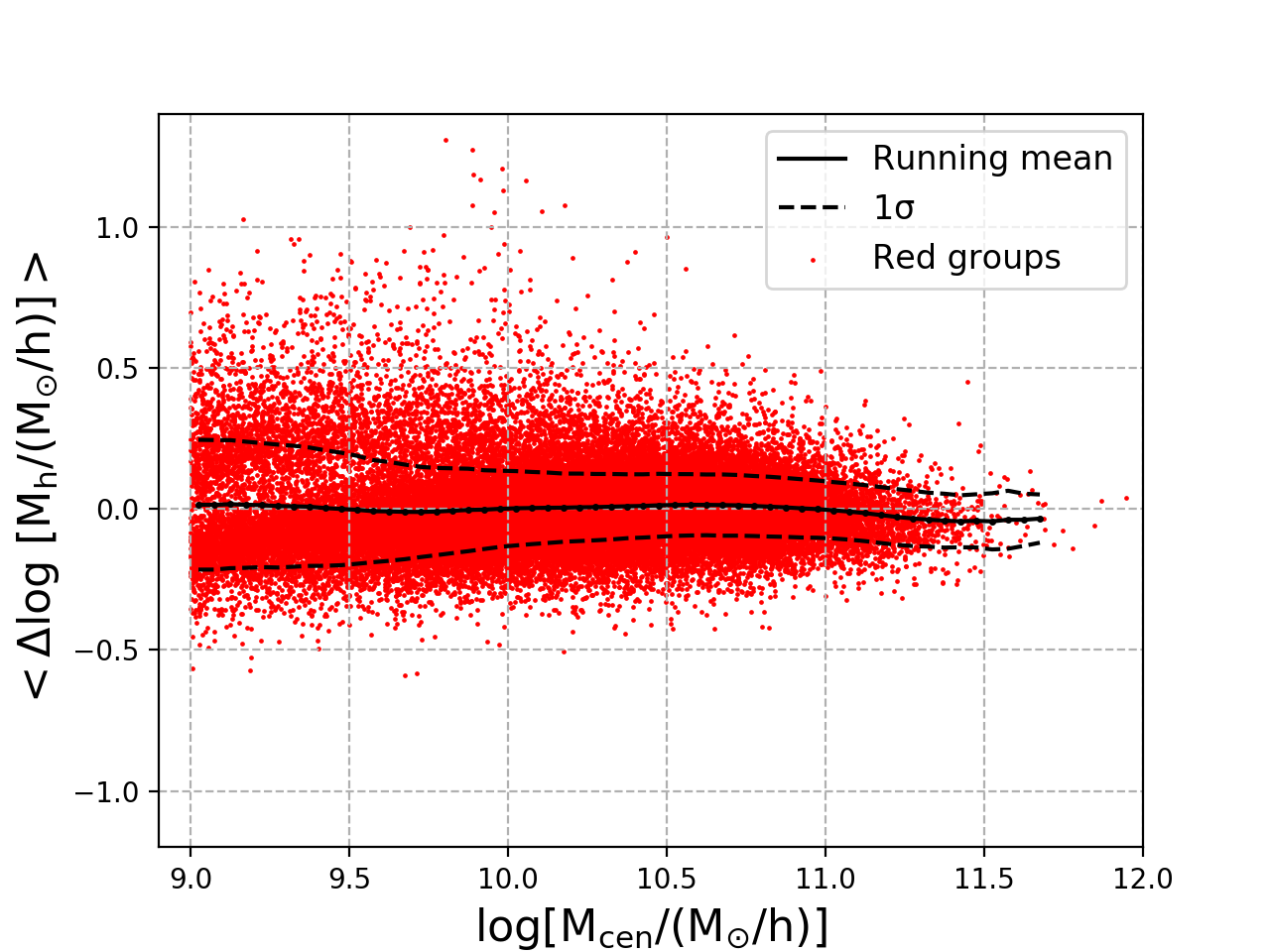}
\includegraphics[width=1\columnwidth]{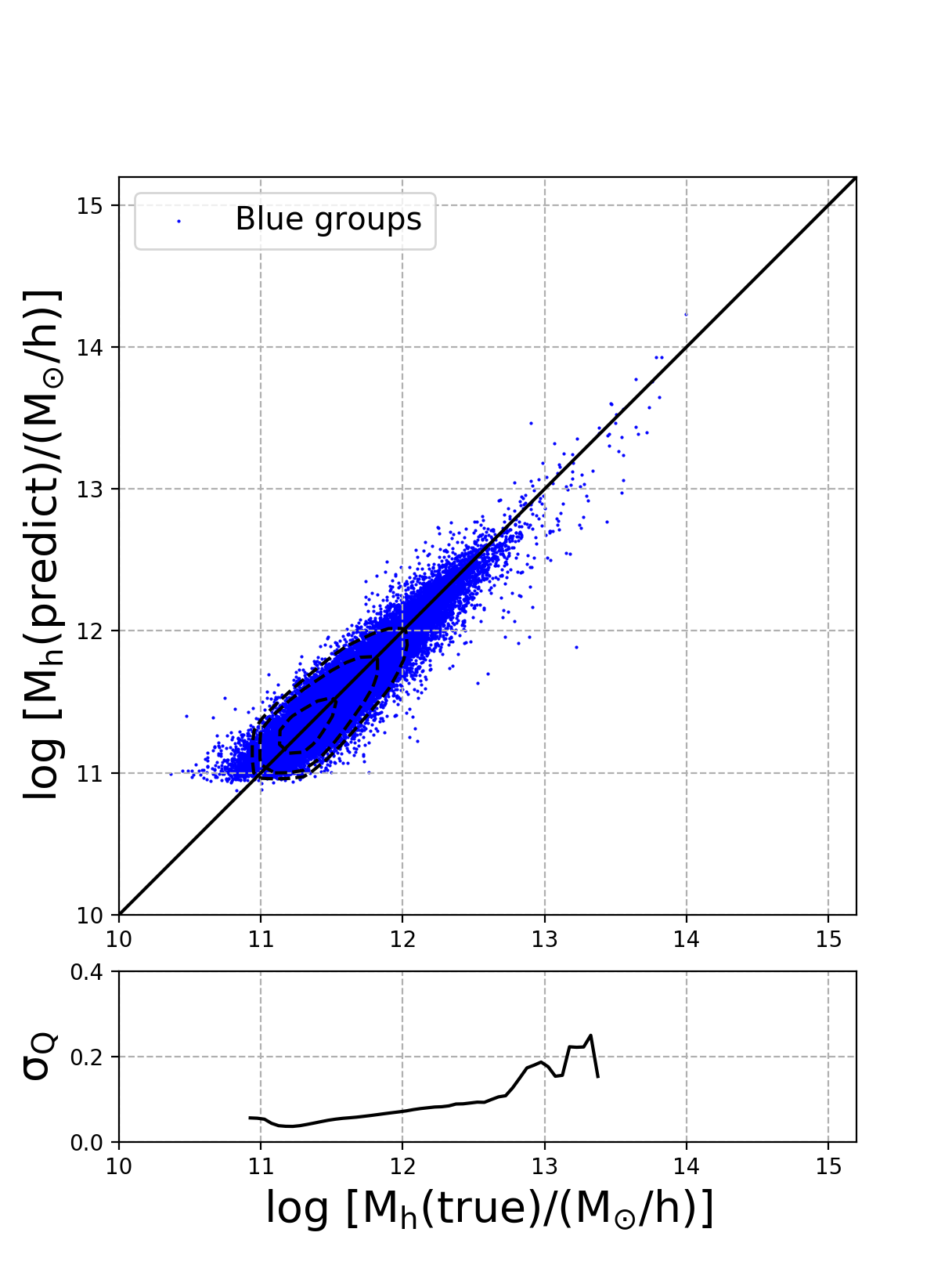}
\includegraphics[width=1\columnwidth]{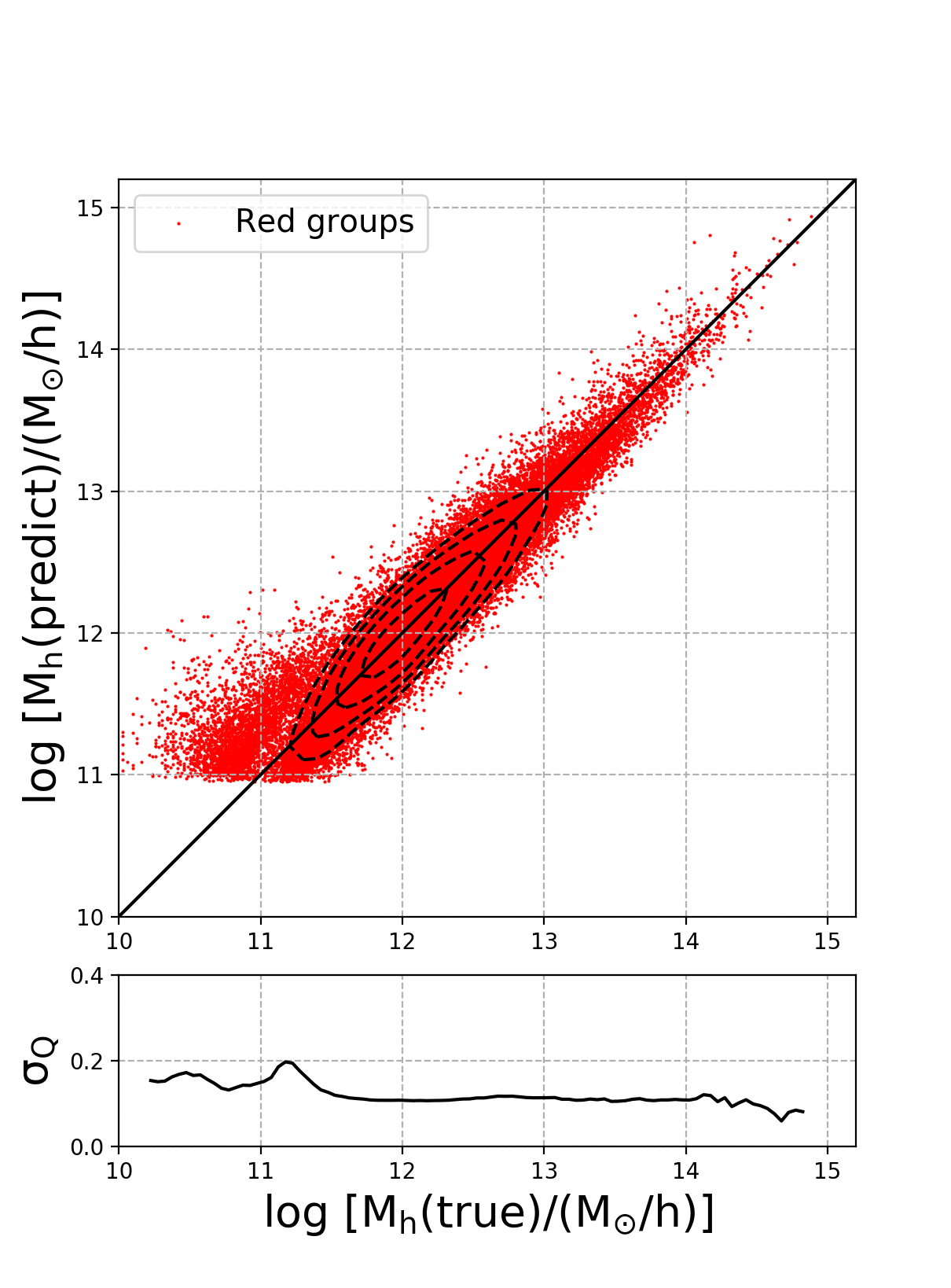}
\caption{Top panels: residuals between the halo mass predicted by the empirical formulae and the true halo mass as a function of the stellar mass of the central galaxy for blue (left) and red (right) groups in the random sample. The running mean and the $1\sigma$ dispersion of residuals are shown as solid and dashed lines. The mean of residuals is almost zero in both panels and is independent of the stellar mass of centrals. The correlation is tighter for blue groups than for red groups, with a smaller $\sigma$ for blue groups. Bottom panels: comparison between the true and predicted halo mass given by the empirical formulae in the random sample. Dashed contour lines show the number distributions of blue and red groups. The bottom subpanels below show $\sigma_Q$ as a function of true halo mass, with a value around 0.1 dex. The MSE is 0.00768 for blue groups and 0.0385 for red groups.\label{fig_5}}
\end{figure*}

We then fit the halo mass with the three key variables suggested by RF regressor. For each variable, we try different forms including logarithm, exponential, and polynomial (up to an order of 3) terms. The combination of them may better fit the data than the simple linear forms. We use the Schwarz information criterion (SIC) as an evaluation of the goodness-of-fit for different models. SIC is used instead of MSE because SIC has a quite strict punishment on the incorporation of additional variables. After searching for various possible models, we obtain the following models, which yield the minimum SIC values. We also perform the $t$ test for the coefficient of each variable and remove the variables whose coefficients are not significant at the 99.9\% level ($p > 0.001$):

\begin{equation}
\label{eq_blue}
\begin{aligned}
{\rm log\ M_h}(blue)&=-15.5319+0.0369\ ({\rm log\ M_{tot}})^3\\
&-0.9445\ ({\rm log\ M_{tot}})^2+8.4697\ {\rm log\ M_{tot}}\\
&-0.0005\ {\rm SFR_n}^3-0.0059\ {\rm SFR_n}^2\\
&+0.0314\ {\rm SFR_n}-0.0791\ \rm (NUV-r)_n\\
& \ \rm R^2=0.956, SIC=0.0077
\end{aligned}
\end{equation}

\begin{equation}
\label{eq_red}
\begin{aligned}
{\rm log\ M_h}(red)&=-21.9131+0.0227\ ({\rm log\ M_{tot}})^3\\
& -0.7279\ ({\rm log\ M_{tot}})^2 +8.3766\ {\rm log\ M_{tot}}\\ 
& +0.0400\ \rm Age_n^2+0.1516\ \rm Age_n\\
& +0.3230\ {\rm log\ Richness}\\
& \ \rm R^2=0.929, SIC=0.0385
\end{aligned}
\end{equation}
where the units for halo mass ($M_{\rm h}$) and total stellar mass ($M_{\rm tot}$) are $M_{\odot}/h$. The renormalized quantities $\rm SFR_n$, $\rm (NUV-r)_n$ and $\rm Age_n$ are dimensionless. Richness is in units of number of galaxies within a given group. The original units for $\rm SFR$ and $\rm Age$ are $log\ M_{\odot}yr^{-1}$ and $Gyr$, respectively.

To make a direct comparison with the results shown in Figure \ref{fig_3} where only test samples are used, we apply the above the empirical formulae to the same test samples, and the results are shown in Figure \ref{fig_5}. The top panels of Figure \ref{fig_5} show $\Delta {\rm log}\ M_{\rm h}$, the residual between the halo mass predicted by equation \ref{eq_blue} and \ref{eq_red} and the true halo mass as a function of the stellar mass of the central galaxy in the random sample. The MSE is 0.00768 for blue groups and 0.0385 for red groups. The mean of residuals is almost zero in both panels, suggesting the scatter is independent of the stellar mass of centrals. This demonstrates that our fitting formulae can be applied to groups spanning a relatively large mass range with little systematic difference. On average, the blue groups have a smaller standard deviation than the red groups. By comparing the predicted halo mass with the true halo mass in the test samples using the two formulae (bottom panels of Figure \ref{fig_5}), we find the $\sigma_Q$ errors in halo mass are reduced by about 50\% from the usual AM approach, as shown in Figure \ref{fig_4}. Although the errors seem comparable with the RF regressor (Figure \ref{fig_3}), the MSEs yielded by the empirical formulae are evidently larger than that of the RF regressor. On the other hand, these simple analytical formulae provide a convenient way to assign halo mass to galaxy groups. 

It should be noted that the total stellar mass of the group and the group richness as in equation \ref{eq_blue} and \ref{eq_red} depend on the sample selection. In the above analysis, only galaxies with stellar mass greater than $10^{9}M_{\odot}/h$ are selected and contribute to the group total stellar mass and richness. A different sample selection will obviously produce different values of $M_{\rm tot}$ and richness. Therefore, the coefficient of each parameter in equation \ref{eq_blue} and \ref{eq_red} must be recalibrated and modified if a different sample selection is used.

Below, we repeat the above fitting process by using a different sample selection of $M_*>10^{10}M_{\odot}/h$.  We use the OLS model with the same analytic forms of equation \ref{eq_blue} and \ref{eq_red} to fit the new sample ($M_*>10^{10}M_{\odot}/h$), and the new formulae are shown as below:

\begin{equation}
\label{eq_blue10}
\begin{aligned}
{\rm log\ M_h}(blue)&=-121.0245+0.1145\ ({\rm log\ M_{tot}})^3\\
&-3.5268\ ({\rm log\ M_{tot}})^2+37.0734\ {\rm log\ M_{tot}}\\
&-0.0003\ {\rm SFR_n}^3-0.0104\ {\rm SFR_n}^2\\
&-0.0214\ {\rm SFR_n}-0.0865\ \rm (NUV-r)_n\\
& \ \rm R^2=0.904, SIC=0.0102
\end{aligned}
\end{equation}

\begin{equation}
\label{eq_red10}
\begin{aligned}
{\rm log\ M_h}(red)&=-24.1263-0.0713\ ({\rm log\ M_{tot}})^3\\
& +2.4413\ ({\rm log\ M_{tot}})^2 -26.9903\ {\rm log\ M_{tot}}\\ 
& +0.0309\ \rm Age_n^2+0.1999\ \rm Age_n\\
& +0.1998\ {\rm log\ Richness}\\
& \ \rm R^2=0.921, SIC=0.0308
\end{aligned}
\end{equation}
where the units for halo mass ($M_{\rm h}$) and total stellar mass ($M_{\rm tot}$) are $M_{\odot}/h$. The renormalized quantities $\rm SFR_n$, $\rm (NUV-r)_n$ and $\rm Age_n$ are dimensionless. Richness is in units of number of galaxies within a given group. The original units for $\rm SFR$ and $\rm Age$ are $log\ M_{\odot}yr^{-1}$ and $Gyr$, respectively.

\begin{figure*}
\centering
\includegraphics[width=1\columnwidth]{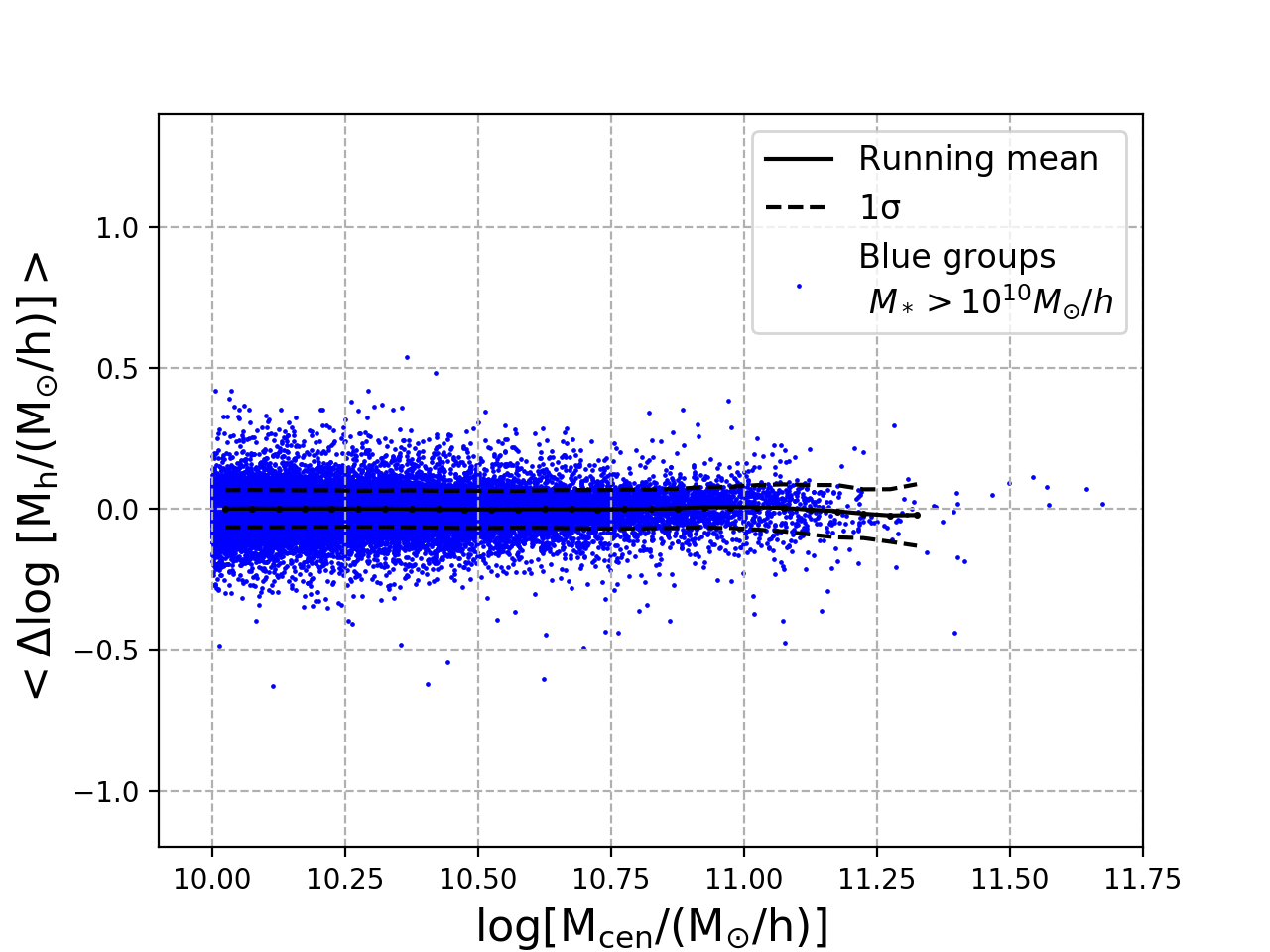}
\centering
\includegraphics[width=1\columnwidth]{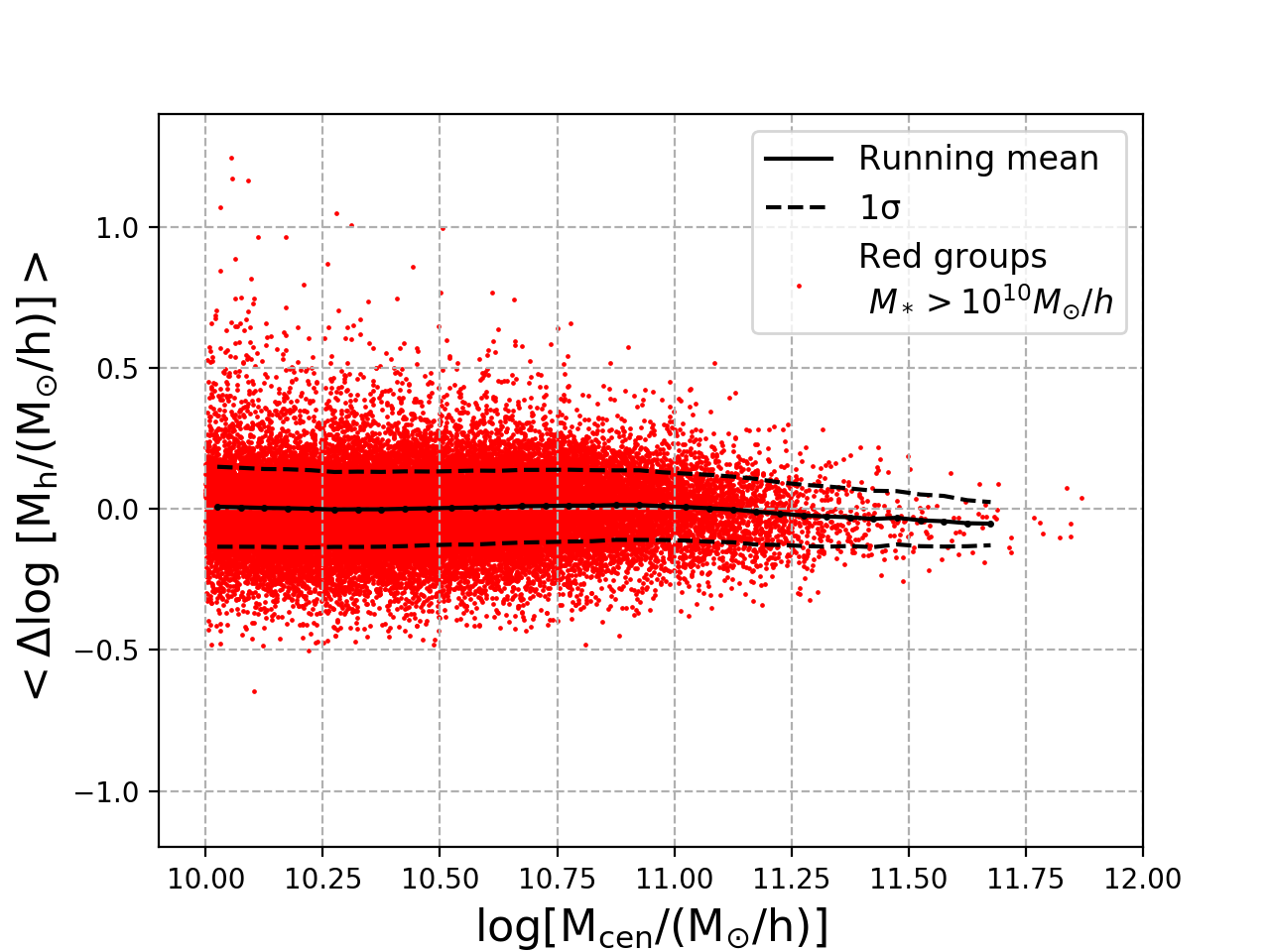}
\includegraphics[width=1\columnwidth]{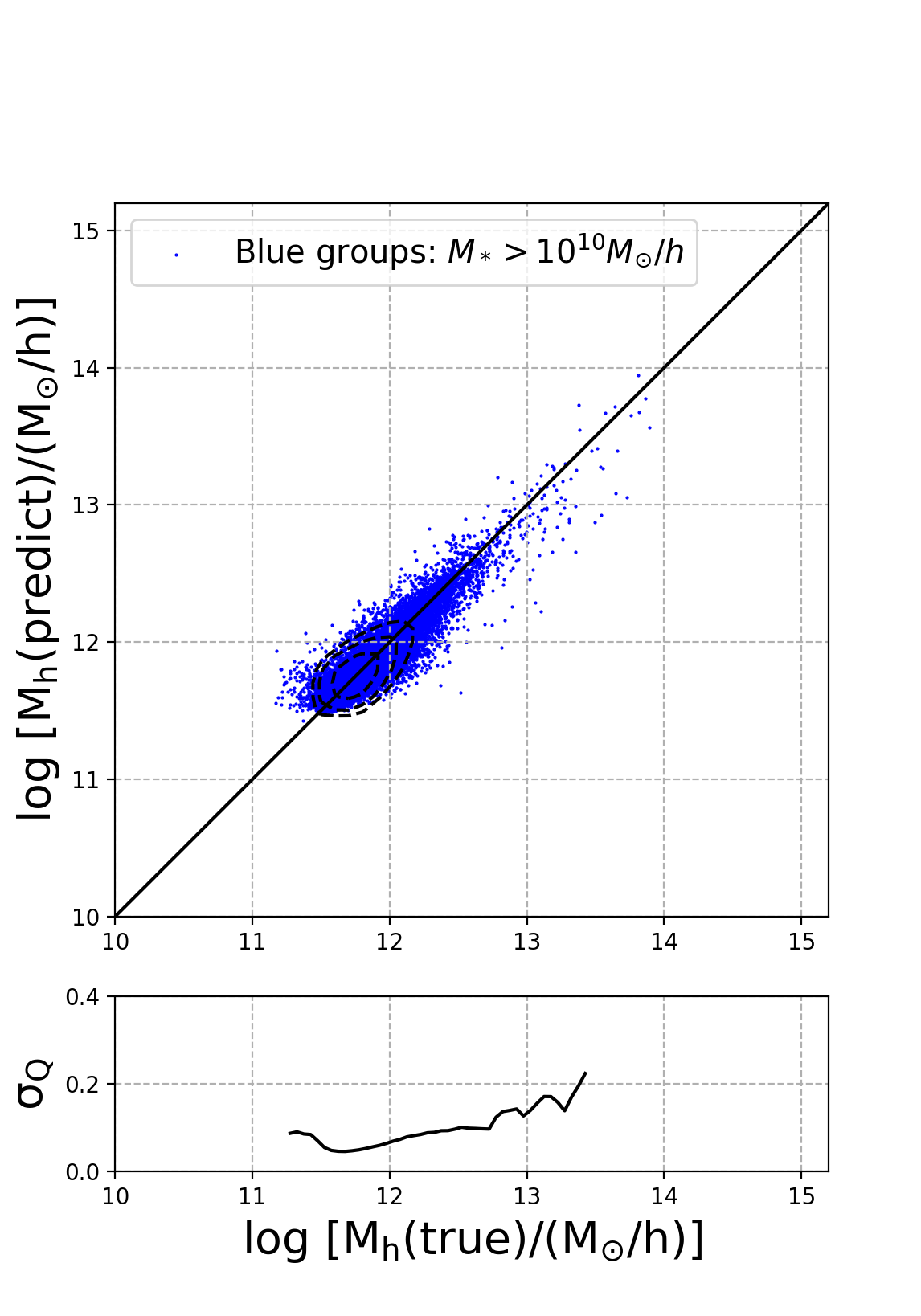}
\includegraphics[width=1\columnwidth]{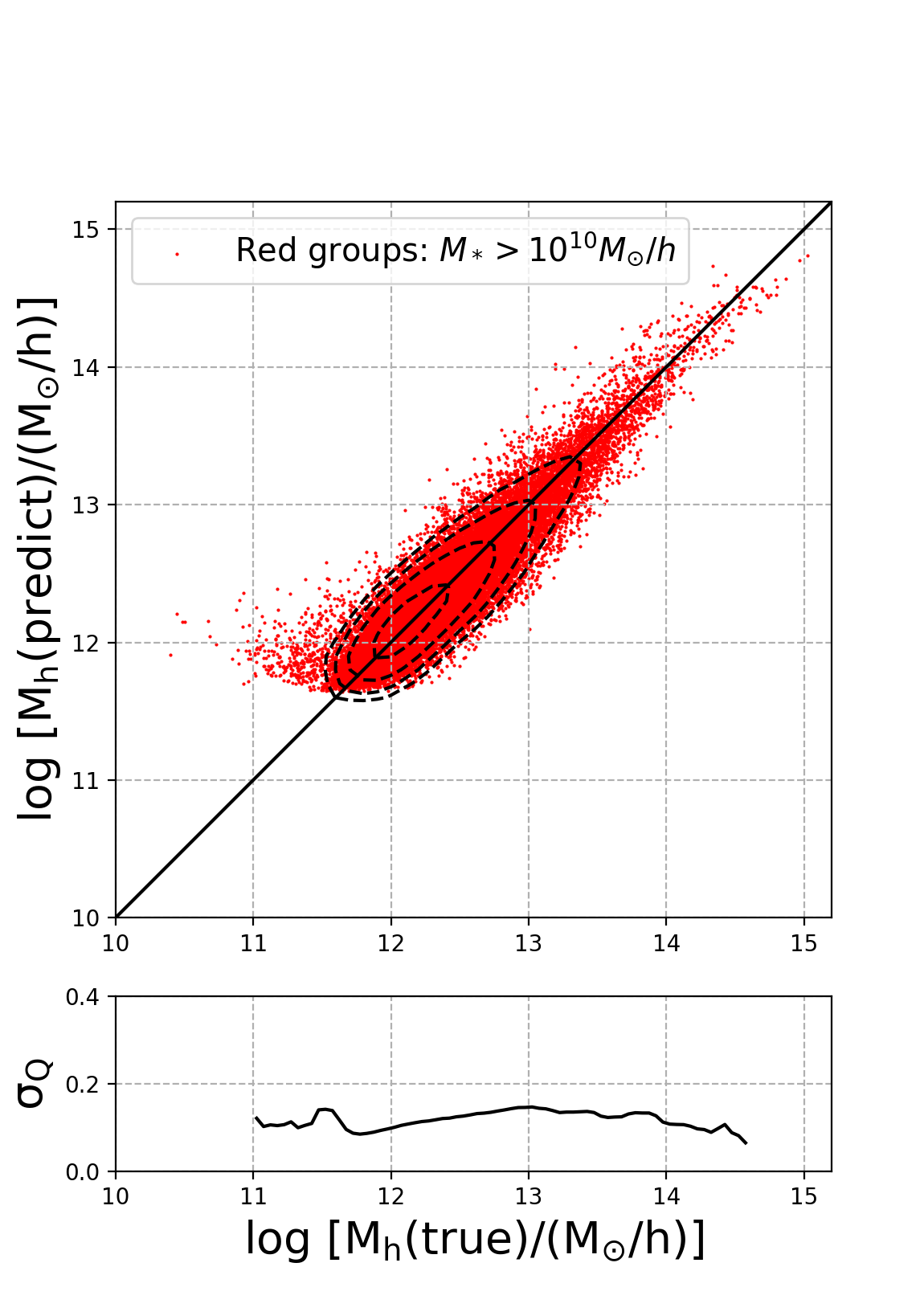}
\caption{As in Figure 5,  but for a different sample selection: only galaxies with a stellar mass greater than $10^{10}M_{\odot}/h$ are selected. \label{fig_6}}
\end{figure*}

As in Figure \ref{fig_5}, the comparisons between the true halo mass and predicted halo mass are shown in Figure \ref{fig_6}. The trends of residuals are similar to those shown in Figure \ref{fig_5}, but with a slightly larger $\sigma_Q$. Also, the minimum halo mass that can be recovered by the fitting formulae is larger for the sample with $M_*>10^{10}M_{\odot}/h$. This is expected because when using a lower mass selection (e.g. $M_*>10^{9}M_{\odot}/h$) more galaxies will be included in the sample. This will produce on average larger $M_{\rm tot}$ and larger richness, thus leading to a more accurate predicted halo mass. However, a lower mass selection will require deeper and longer observations.

All input parameters in the above equations are observable quantities. As discussed in section \ref{subsec_group_properties}, there is always a systematic difference in the absolute mean values (e.g. stellar mass, SFR, color) between model predictions and observations. Similarly, the scatters around the mean values are also different. To reduce any bias introduced by the systematic difference in both the mean value and scatter around the mean, we have used the renormalized dimensionless forms $\rm P_n$, as in equation \ref{equ_1}, in all our analyses. Therefore, in principle, equation \ref{eq_blue} - \ref{eq_red10} can be applied directly to real observational data. 

Nevertheless, there are still complications and caveats. Apparently, the renormalized dimensionless form will reduce but cannot fully remove all systematic differences between observations and simulations, in both the mean value and scatter around the mean. The mean and scatter can also vary in different observations and surveys. Therefore, ideally one should first compare the values observed in a certain survey, including stellar mass, color, SFR, and stellar age, with the L-GALAXIES predictions. If there are no significant differences in the mean and scatter of these values, the above analytic equations can be applied to  derive the halo mass safely. In our future work, we will apply our method to surveys such as SDSS, GAMA, and COSMOS and etc. to derive the group halo mass. We will also apply our approach to the latest hydrodynamical simulations like EAGLE (\citealt{Cra 15}; \citealt{Sch 15}) and IllustrisTNG (\citealt{Nel 19}), to see if we will get consistent results. These latest hydrosimulations may produce more accurate properties such as SFR and metallicity and hence may provide a more accurate description of the galaxy$-$halo connection. On the other hand, SAMs like L-GALAXIES still produce more accurate SMFs than hydrosimulations, which indicates a more accurate SHMR. Also, SAMs usually have a much larger volume than the hydro-simulations, leading to larger training samples and less cosmic variance. This is also the reason why we start our analysis with L-GALAXIES in the first place.

Another important uncertainty comes from the group finder in observations. Group finders are designed to identify galaxy group members in the same dark matter halos based on their spatial distributions (e.g. using friends-of-friends technique). However, none of these group finders can fully recover the true group membership for each galaxy. The purity and completeness of all the recovered groups can never reach 100\%. For instance, for any group catalogs, even if carefully calibrated against mock catalogs in which the underlying dark matter distribution is known, overfragmentation and overmerging of groups still happen. This leads to misclassification of satellites and centrals and produces wrong group richness and total stellar mass in the group; that is, the input quantities used in our equation \ref{eq_blue} and \ref{eq_red} could be wrong. Apparently, this will be a common issue for all studies related to galaxy groups. We will try to quantify this effect in our future study.

\section{Summary}
\label{sec_sum}
In this paper, we have investigated the key relationships between the group halo mass and various observable galaxy group properties using the semianalytical galaxy formation models L-GALAXIES. We first propose a simple scenario (illustrated in Figure \ref{fig_2}), which describes the evolution of the central galaxies and their host dark matter halos. Star formation quenching is one of the key processes in this scenario, which leads to the different assembly histories of blue groups (group with a blue central) and red groups (group with a red central). From this simple scenario, we speculated about the driving factors that should strongly correlate with the group halo mass. We then apply ML algorithm RF regressor to blue groups and red groups separately, to explore these nonlinear and nonorthogonal multicorrelations and to verify the scenario as proposed above. Remarkably, the results given by RF regressor are fully consistent with the prediction from our simple scenario. As a consequence, the group halo mass can be more accurately determined from observable galaxy properties by the RF regressor.

The main results of the paper are summarized as follows:

(1) The total stellar mass of a group is expected to correlate most strongly with the group halo mass, because the total stellar mass is the best indicator of both the overall SFH of all group members and the merging history of the halo.  

As illustrated in Figure \ref{fig_2}, for blue groups, because both the central galaxy and its host halo will continue to grow their masses simultaneously, a relatively tight relation between the stellar mass of the central and its host halo mass is expected. Therefore, in addition to the total stellar mass of the group, the properties that indicate the SFH of the centrals, such as SFR and color of the centrals, should also correlate strongly with the group halo mass.

For red groups, when the central is quenched at some point, its stellar mass remains about constant unless additional stellar mass is accreted through subsequent mergers, while its halo continues to grow by merging smaller halos. In other words, the growth of the halo is decoupled from the growth of the central.

Therefore, in addition to the total stellar mass of the group, the properties that can indicate the quenching epoch of the centrals (e.g. stellar age of the central) and the halo growth history (e.g. group richness) should also correlate strongly with group halo mass. The distinct evolutions of blue and red groups, due to the quenching of the centrals, require that we must treat them separately in our analysis.

(2) By using the RF regressor,  among the various group properties explored, we find that the total stellar mass of the group is the most important parameter for both blue and red groups, followed by the SFR and NUV$-$r color of the central galaxy for blue groups and group richness and stellar age of the central galaxy for red groups. This is fully consistent with the simple scenario proposed above and hence provides strong support for it. 

Since the ML algorithm can also quantify the correlation between various observable galaxy properties and group halo mass, in return, the group halo mass can be more accurately predicted from observable galaxy properties. Compared to the traditional AM approach, the standard errors in the halo mass predicted by the RF regressor have been reduced by about 50\%.

(3) The blue and red groups are classified according to the color$-$color diagram of the central galaxies. Although the color of the centrals has already been included as an input parameter in the RF regressor, running RF regressor separately for blue and red groups can produce more accurate halo masses. RF regressor is a powerful ML algorithm, yet it failed to capture the quenching process accurately by itself. Therefore, by taking quenching into account (i.e., differentiate between blue and red groups) when performing RF regressor, we have improved its performance and produced a more accurate prediction of the halo mass. Another potentially important hidden process is merger. If we could quantify the composition of stellar mass for a given central galaxy (e.g., how much of its stellar mass is from in situ star formation and how much is from mergers), we should be able to further improve the accuracy of the halo mass prediction by including it as a new input parameter. We will explore this in our subsequent work.

This implies that a better understanding of the underlying physics, in particular those hidden deep correlations between multiple variables, will help to improve the performance of the ML algorithms.

(4) Similar to other ML algorithms, RF regressor does not give an explicit form of the relation between group halo mass and group properties. We hence regress the halo mass on the key variables identified by RF regressor, and we derive the empirical relations that can be used to determine the halo mass analytically. Since the total stellar mass of the group and group richness that are used in these relations as input parameters depend on the sample selection, we proposed equations \ref{eq_blue} and \ref{eq_red} for a sample of galaxies with stellar mass greater than $M_*>10^{9}M_{\odot}/h$; and equations \ref{eq_blue10} and \ref{eq_red10} for a sample of galaxies with stellar mass greater than $M_*>10^{10}M_{\odot}/h$. These simple analytical formulae provide a convenient way to assign halo mass to galaxy groups from observable group properties, with accuracy comparable to those determined directly from the RF regressor.

In our future work, we will include more observable properties of the galaxy groups, for instance, the structure, morphology, and dynamics of the group members. As mentioned at the end of section \ref{subsec_feature_importance}, another potentially important hidden process (besides quenching) is merger. If we could quantify it for a given central galaxy (e.g., how much of its stellar mass is from in-situ star formation and how much is from mergers), we should be able to further improve the accuracy of the halo mass prediction.  We will also test our approach with the latest hydrodynamical simulations like EAGLE (\citealt{Cra 15}; \citealt{Sch 15}) and IllustrisTNG (\citealt{Nel 19}) to see if we will get consistent results.  

Then we will apply the RF regressor to surveys such as SDSS, GAMA, and COSMOS to derive more accurate halo masses, which will enable more accurate investigations of the galaxy$-$halo connection and many other important related issues, including galactic conformity and the effect of halo assembly bias on galaxy assembly.

\acknowledgments
We are grateful to Frank C. van den Bosch, Zheng Zheng, and Qi Guo for the productive discussions and useful comments. We thank the anonymous referee for useful comments. We are particularly grateful to the L-GALAXIES project for making the data public. We also thank Aobo Li for helping polish the text. This work is supported by the National Natural Science Foundation of China grant No. 11773001 and National Key R\&D Program of China grant 2016YFA0400702. J.S. acknowledges the support by the Peking University Boya Fellowship. X.K. acknowledges the support by the National Key R\&D Program of China grant 2015CB857004 and 2017YFA0402600, and the National Natural Science Foundation of China grant No. 11320101002, No. 11421303, and No. 11433005. K.G. acknowledges the support from the Beijing Natural Science Foundation (Youth program) under grant No. 1184015.

 \end{document}